\documentclass[useAMS,usenatbib]{mn2e}

\usepackage{graphicx}
\usepackage{subfigure}
\usepackage{color}
\usepackage[authoryear]{natbib}
\usepackage{names}

\def\ie{{\it i.e.} }

\def\Herschel{{\it Herschel } }
\def\Herschelnospace{{\it Herschel} }

\def\kcrb{$\kappa$ CrB }
\def\kcrbdot{$\kappa$ CrB. }
\def\kcrbcomma{$\kappa$ CrB, }
\def\kcrbs{$\kappa$ CrB's }

\newcommand\blfootnote[1]{%
  \begingroup
  \renewcommand\thefootnote{}\footnote{#1}%
  \addtocounter{footnote}{-1}%
  \endgroup
}

\begin{document}
\title[]{   Spatially Resolved Images of Dust Belt(s) Around the Planet-hosting Subgiant \kcrb }

  \author[A. Bonsor et al.]{Amy Bonsor$^1$\thanks{Email:amy.bonsor@gmail.com}, Grant M. Kennedy$^2$, Justin R. Crepp$^3$,  John A. Johnson$^4$, \newauthor Mark C. Wyatt $^2$, Bruce Sibthorpe$^5$ and Kate Y. L. Su$^6$\\
$^1$UJF-Grenoble 1 / CNRS-INSU, Institut de Planétologie et d'Astrophysique de Grenoble (IPAG) UMR 5274, Grenoble, F-38041, France  \\     
$^2$Institute of Astronomy, University of Cambridge, Madingley Road, Cambridge CB3 OHA, UK \\
$^3$Department of Physics, University of Notre Dame, 225 Nieuwland Science Hall, Notre Dame, IN 46556, USA \\
$^4$Department of Astronomy, California Institute of Technology, 1200 E. California Blvd., Pasadena, CA 91125, USA \\
$^5$SRON Netherlands Institute for Space Research, Zernike Building, P.O. Box 800, 9700 AV Groningen, The Netherlands \\
$^6$ Steward Observatory, University of Arizona, 933 N Cherry Ave., Tucson, AZ 85721 
}

   \date{Accepted 2013 February 24.  Received 2013 February 22; in original form 2012 December 20}

\maketitle

\begin{abstract}

We present \Herschelnospace$^1$ spatially resolved images of the debris disc orbiting the subgiant \kcrbdot Not only are these the first resolved images of a debris disc orbiting a subgiant, but \kcrb is a rare example of an intermediate mass star where a detailed study of the structure of the planetary system can be made, including both planets and planetesimal belt(s). The only way to discover planets around such stars using the radial velocity technique is to observe `retired' A stars, which are cooler and slower rotators compared to their main-sequence counterparts. A planetary companion has already been detected orbiting the subgiant \kcrbcomma with revised parameters of $m \sin i = 2.1M_J$ and $a_{pl}=2.8$AU  \citep{Johnson08}. We present additional Keck I HIRES radial velocity measurements that provide evidence for a second planetary companion, alongside Keck II AO imaging that places an upper limit on the mass of this companion. Modelling of our \Herschel images shows that the dust is broadly distributed, but cannot distinguish between a single wide belt (from 20 to 220AU) or two narrow dust belts (at around 40 and 165AU). Given the existence of a second planetary companion beyond $\sim$3AU it is possible that the absence of dust within $\sim20$AU is caused by dynamical depletion, although the observations are not inconsistent with depletion of these regions by collisional erosion, which occurs at higher rates closer to the star. 


\end{abstract}

\section{Introduction}
\label{sec:intro}

Our knowledge\blfootnote{$^1$ \Herschel in an ESA space observatory with science instruments
provided by European-led Principal Investigator consortia and with important
participation by NASA } and understanding of exo-planetary systems is growing rapidly. Since the first detection of a Kuiper-like, planetesimal belt in 1984 (Vega, \cite{Aumann1984}), the first planet detection around a pulsar in 1992 \citep{pulsarplanets} and a close-in Jupiter-mass planet around a main-sequence star in 1995 \citep{Mayor1995}, the field has exploded. There are now hundreds of systems with planet or debris disc detections. There is a great deal to be learnt from the growing number of stars where both planets and debris discs have been detected.

Current planet detection techniques are limited to specific regions of the parameter space. For example, radial velocity observations are limited to the inner regions of planetary systems, whilst direct imaging is limited to the outer regions. This means that in order to fully characterise a planetary system, it is beneficial to have simultaneous access to data from different detection techniques. Radial velocity observations of A stars on the main-sequence are prohibited due to high jitter levels and rotationally broadened absorption lines \citep{Galland05, Lagrange09}, however, there are now a growing number of detections of planets around `retired' A stars, now on the subgiant or giant branch \citep[e.g][]{Johnson06,Johnson07,Bowler2010, Sato2010}. These provide some key insights into the potential differences between the planetary population around intermediate mass stars, that otherwise can only be probed by direct imaging of planets around main-sequence A stars \citep[e.g.][]{hr8799detection08, fomb2008}. For example, \cite{Bowler2010} and \cite{Johnson_planetpopulation} found an increased incidence of giant planets around stars of higher stellar mass, as predicted by planet formation models \citep{Kennedy_Kenyon2008}. 

There are a growing number of sun-like stars with both planet and debris disc detections \citep[e.g][]{Wyatt61Vir, Lestrade2012, Liseau2010}. Such systems provide key insights into the structure of exo-planetary systems and the interactions between planetesimal belts and planets. Resolved debris discs often display a variety of features that can be associated with the presence of planets, amongst others, warps, spirals, brightness asymmetries, clumps and offsets \citep[e.g.][]{Augereau01, Moerchen10, Wyatt99}. Gaps between multiple planetesimal belts could potentially be cleared by unseen planetary companions, whilst planets may commonly sculpt the inner or outer edges of planetesimal belts \citep[e.g.][]{Hr8799su, chiang_fom, Churcher10, Lagrange2012}. Despite the ubiquity of debris discs around main-sequence A stars \citep{wyatt07, Booth2012} and direct imaging of a handful of distant planets \citep{hr8799detection08, Lagrange2010}, the inner planetary systems remain poorly constrained due to aforementioned problems with radial velocity measurements. The best way to learn about the inner planetary systems of intermediate mass stars is therefore to observe `retired' A stars. Very little, however, is known about debris discs around such `retired' A stars. Such knowledge could act as a further window onto the structure of planetary systems around intermediate mass stars, critical to furthering our understanding of planetary systems in general.

In this work we present \Herschel images of a debris disc around the subgiant $\kappa$ Coronae Borealis (\kcrbcomma HD 142091, HR 5901, HIP 77655) and resolve excess emission in the far-infrared. \kcrb is a K-type subgiant near the base of the giant branch with a mass of 1.8$M_{\odot}$  at a distance of 31.1pc \citep{Johnson08}\footnote{Calculated using the stellar models of \cite{Girardi2002}}. \kcrb is significantly cooler than the average main sequence A star, but not significantly more luminous, with a luminosity of $12.3L_\odot$ and age of 2.5 Gyr \citep{Johnson08}. Radial velocity monitoring of \kcrb using the Lick observatories \citep{Johnson08} found evidence for a planetary companion. The best fit to the radial velocity variations find a $m \sin i = 2.1M_J$ planet at $2.8\pm0.1$AU, with an eccentricity of $0.125 \pm 0.049$ \footnote{Updated from the $m \sin i = 1.8M_J$, 2.7AU and $e=0.146\pm 0.08$ values quoted in \cite{Johnson08}.}. We present far-infrared \Herschel observations of this source that find and resolve excess emission, alongside follow-up radial velocity measurements that suggest the presence of a second companion and direct imaging attempts with Keck that constrain the potential orbital parameters of this companion.

We start by presenting the observations in \S\ref{sec:obs}, followed by the basic results determined from these observations in \S\ref{sec:results}. Detailed modelling of the \Herschel images is presented in \S\ref{sec:models}, followed by a discussion of the structure of the \kcrb planetary system in \S\ref{sec:discussion} and our conclusions are made in \S\ref{sec:conclusions}.


\section{Observations}
\label{sec:obs}

\subsection{Keck Radial Velocity Monitoring}
\begin{figure}
\includegraphics[width=0.48\textwidth]{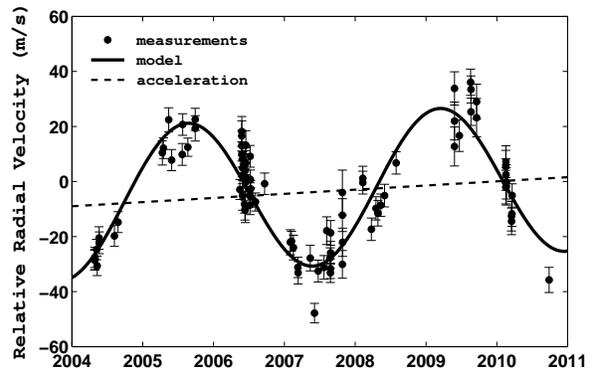}

\caption{ Radial velocity monitoring of \kcrb over 8.09years, showing the new orbital fit for \kcrb b and a Doppler acceleration, that provides evidence for a second companion.  }
\label{fig:RV}
\end{figure}

We monitored \kcrb at Lick observatory from 2004 to 2009, and at Keck observatory from 2010 until present, to search for companions to stars more massive than the Sun. This monitoring found the $m \sin i =1.8M_J$ companion at 2.7AU \citep{Johnson08} in 2008. Since then, continued monitoring of this star, over a total of 8.09 years, has updated the orbital parameters for \kcrb b (shown in Table~\ref{tab:planet}) and found $m\sin i = 2.1 M_J$, a semi-major axis of $2.8\pm0.1$AU, as well as a Doppler acceleration of $1.51 \pm 0.52$ ms$^{-1}\mathrm{yr}^{-1}$. Such a trend provides good evidence for the presence of a second companion, however, further monitoring is required before the orbit of this companion can be constrained. The radial velocity curve for this target is shown in Fig.~\ref{fig:RV}.

\begin{table}
\begin{tabular}{c| c| c}

Period & P &$  1300 \pm 15$ days \\
Time of pericentre passage &$T_p$ &$ 13899 \pm 160$ JD \\
Eccentricity & e  &$ 0.125 \pm 0.049$\\
Argument of pericentre&$\omega$ &$ 83.1 \pm 29$ deg\\
Velocity semi-amplitude & K  &$27.3 \pm 1.3$ms$^{-1}$ \\
 Acceleration &$\frac{dv}{dt}$ &$ 1.51\pm 0.52 $ms$^{-1}$yr$^{-1}$\\

\end{tabular}
\caption{The new best-fit orbital parameters for \kcrb b derived from the continued radial velocity monitoring at the Lick and Keck observatories. This fit had a reduced chi-squared value of 1.8 and 7 free parameters, namely, period, eccentricity, longitude of periastron, time of periastron passage, global RV off-set, semi-amplitude and acceleration. These are derived using the same bootstrap Monte Carlo method, as described in further detail in \citet{Johnson08}. 
}
\label{tab:planet}
\end{table}


\begin{figure}
\includegraphics[width=0.48\textwidth]{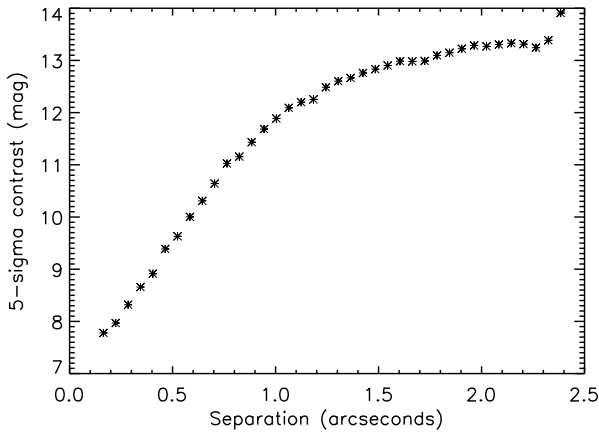}

\caption{ The upper limits on emission from the region surrounding \kcrb determined by the Keck AO imaging. This is converted to upper limits on the companion mass in \S\ref{sec:detlimits}, shown in Fig.~\ref{fig:det_lim}. }
\label{fig:raw}
\end{figure}

\subsection{Adaptive Optics Observations}
\label{sec:AO}
 
Given evidence for the existence of an additional companion in the system with a period at least as long as the observational baseline (8.09years), \kcrb was observed as part of the TRENDS imaging program - a survey dedicated to follow-up high-contrast observations of stars showing long-term Doppler accelerations \citep{Crepp2012}. Using NIRC2 (PI Keith Matthews) and the Keck II adaptive optics (AO) system \citep{Wizinowich2000}, angular differential imaging observations were acquired on May 26, 2010 in an attempt to directly image the outer body responsible for accelerating the star.

A total of 90 frames were recorded using the narrow camera setting. Each frame consisted of a 30 second integration time (60 coadds with 0.5 seconds per coadd), resulting in a total on-source integration time of 2700 seconds. The field of view ($10"$ x $10"$, modulo a bad detector quadrant) 
 was allowed to rotate to help discriminate between residual scattered starlight (quasi-static speckles) and faint candidate companions \citep{Marois2006}. The parallactic angle changed by 49.5 degrees during the course of the observations, allowing us to achieve a close (150 mas) inner-working angle. The airmass ranged from 1.04 to 1.08.

\kcrb is sufficiently bright (H=2.58) that observations were acquired with the Hcont narrow-band filter. The 300 mas diameter coronagraphic spot was used to occult the star. Unocculted frames were also obtained to measure the relative brightness (contrast) between the primary star and any off-axis sources.

Individual frames were processed using standard techniques to flat-field the array, identify and replace hot pixel values, and align and co-add images. We used the locally-optimized combination of images (LOCI) algorithm to improve the effective signal-to-noise ratio of speckle suppressed frames \citep{Lafreniere2007}. We did not detect any candidate companions.

The data, the $5$-$\sigma$ contrasts as a function of angular separation, are shown in Fig.~\ref{fig:raw}. These are later converted to upper mass limits on the second companion in \S\ref{sec:detlimits}.
 
\subsection{The \Herschel Observations}

Observations were performed using the Herschel Photodetector and Array
Camera \& Spectrometer (PACS, \cite{Poglitsch2010}) at 100 and
160$\mu$m, as listed in Table~\ref{tab:obs}.  These observations were performed in
mini scan-map mode with two observations being performed with a $40\deg$
cross-linking angle. Four repeats were used for each observation and
with eight scan legs per repeat.  The total observing time was
approximately 30 minutes.

Data were reduced with the Herschel Interactive Processing Environment
version 7.0 Build 1931 (HIPE, \cite{Ott2010}) using version 32 of the PACS
calibration.  Some data from the telescope turn-around phase (when
scanning above $5\arcsec/s$) were used to minimize the ultimate noise
level.  Maps were then made using the HIPE photProject task to provide
`drizzle' maps \citep{Fruchter_Hook2002} with pixel scales of 1 and 2
arcsec in the 100 and 160\,$\mu$m bands respectively.  The data were
high-pass filtered to mitigate low frequency 1/$f$ noise, using
filtering scales of 66 and 102 arcsec (equivalent to a filter radius
of 16 and 25 PACS frames) in the 100 and 160\,$\mu$m bands
respectively.

The point-spread function (PSF) of the PACS beam includes significant
power on large scales (10\% beyond 1 arcmin).  Consequently, the
filtering performed during the data reduction will reduce the flux
density of a source by $10-20\%$, due to the filter removing the ‘wings’
of the PSF.  For point sources this can be readily accounted for using
correction factors, determined from comparison of bright stars with
known fluxes with the PACS aperture flux. Correction factors of $1.19 \pm 0.05$ and $1.12 \pm 0.05$ at 100 and 160$\mu$m were determined from analysis of the DEBRIS (Disc Emission via a Bias-free Reconnaissance
in the Infrared/Submillimetre) survey \citep[e.g.][]{Matthews2010}.
DEBRIS targets \citep{Kennedy99Her}. This can also be applied to resolved sources when the
source remains similar in scale to the beam Full Width Half Maximum (FWHM).


\begin{table}

\begin{tabular}{l l l l l}
\hline
Target &obsID & Date & Instrument & Duration  \\
\hline
\kcrb & 1342234353 & 15/12/2011 & PACS & 895s \\
\kcrb &1342234354 & 15/12/2011 &  PACS & 895s \\

\hline

\end{tabular}
\caption{ The \Herschel observations. }
\label{tab:obs}
\end{table}


\begin{table}

\begin{tabular}{|c |c |c c|}

\hline
Instrument &Wavelength & Photosphere & Observed \\ 
 & $\mu$m & mJy & mJy\\
\hline
Spitzer &24  & $766 \pm 13$ & 800.1  $\pm 0.1 \pm 8.0$$^a$ \\
Spitzer &70  & $83 \pm 2$  & $426.2^b \pm 6.5 \pm 22.3$$^a$ \\
\Herschel &100 & $42.34\pm 0.69$ & $335\pm  16$mJy\\
\Herschel &160& $16.40 \pm 0.27$& $192\pm 10 $mJy \\

\hline
\end{tabular}
\caption{The photospheric fluxes for \kcrb compared to observed fluxes from both archival Spitzer data \citep{Kalas_proposal} and the \Herschel data presented in this work, including systematic uncertainties. $^a$ For the Spitzer data the first uncertainties quoted are photometry uncertainties estimated based on the data, whilst the second ones include the overall calibration. $^b$ The photometry quoted for MIPS $70\mu$m is based on aperture photometry rather than the usual PSF fitting photometry, as the source is slightly elongated. }
\label{tab:flux}
\end{table}


\begin{figure*}
\includegraphics[width=0.48\textwidth]{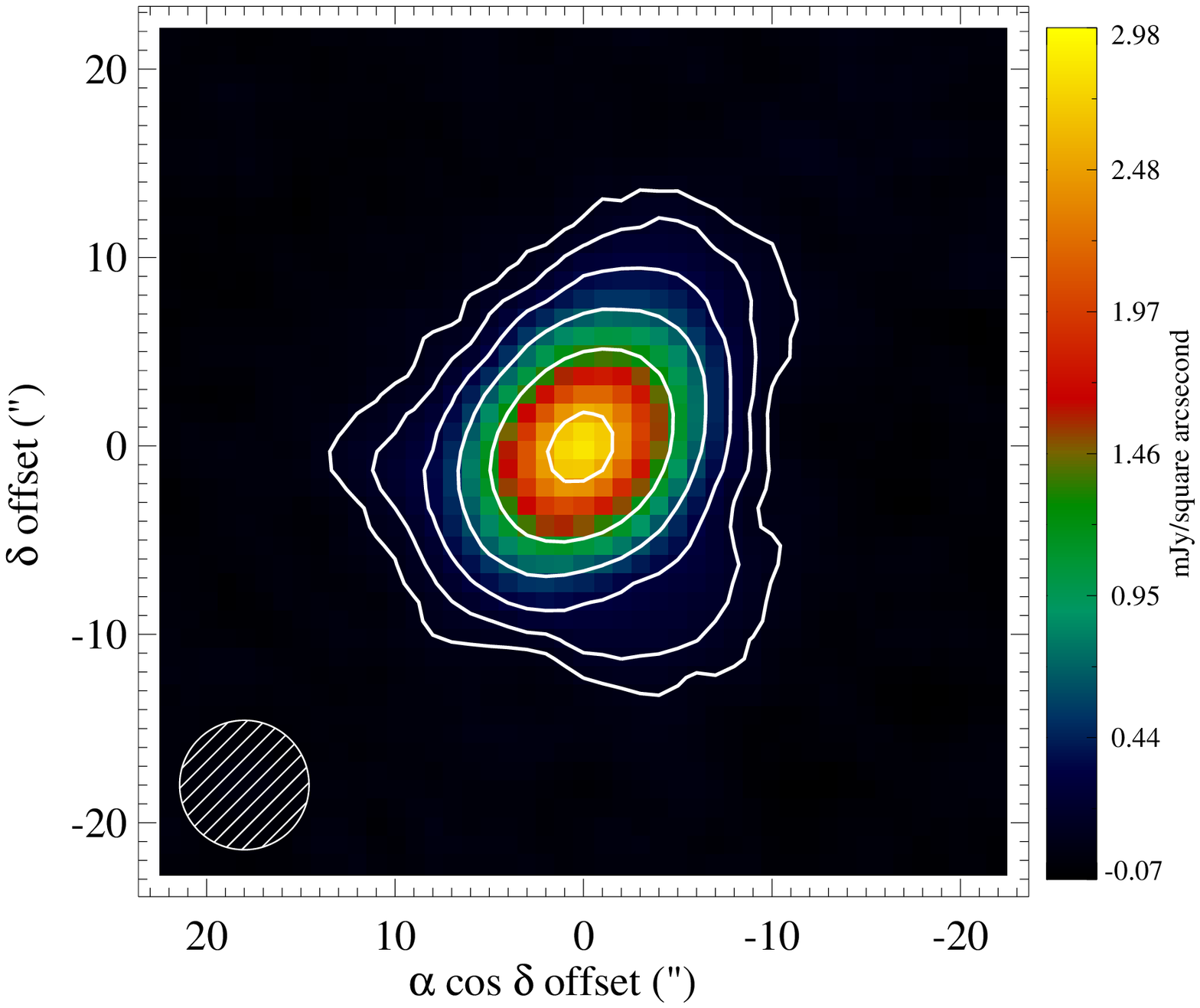}
\includegraphics[width=0.48\textwidth]{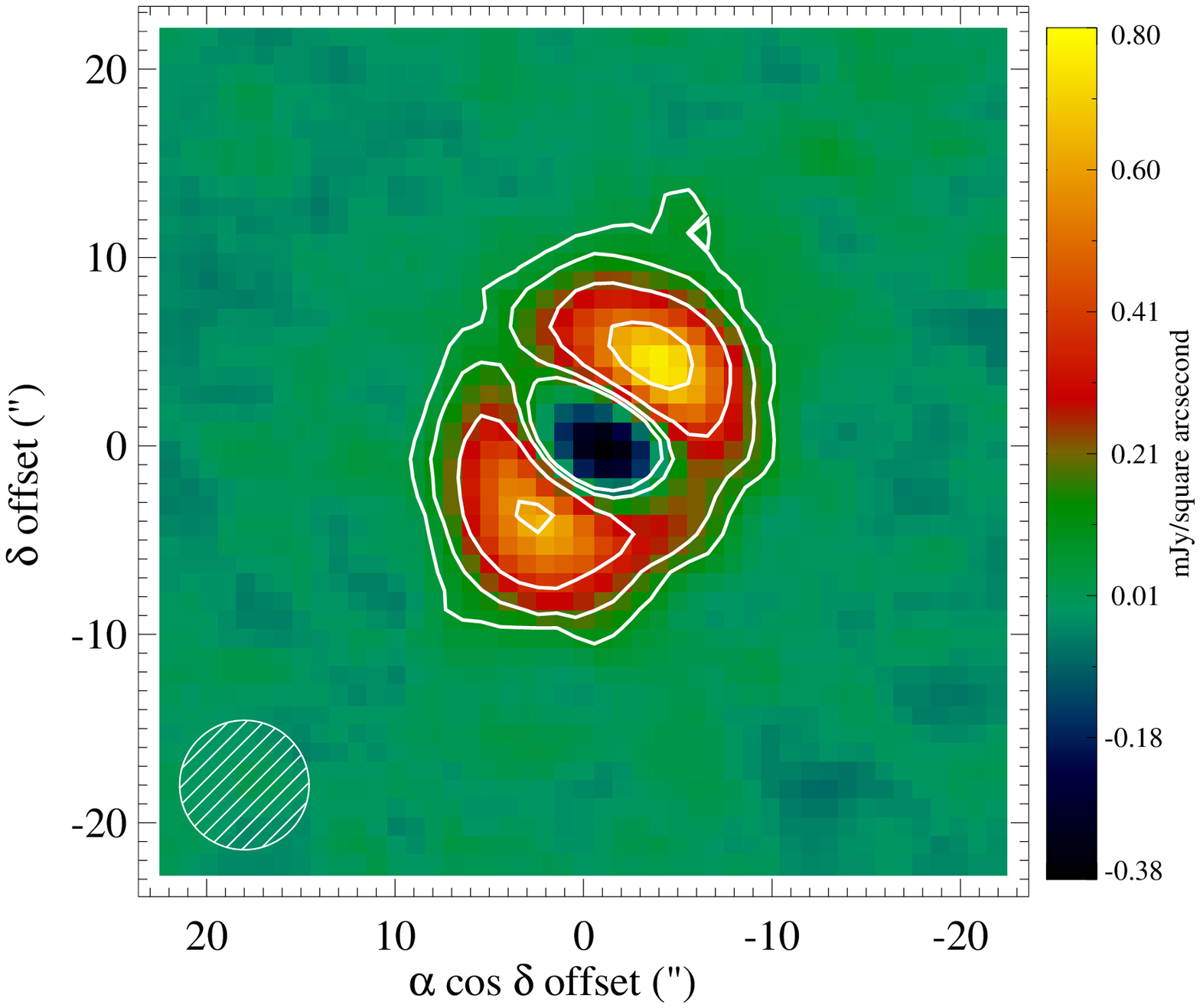}

\caption{ The 100$\mu$m \Herschel PACS observations of \kcrbdot North is up and East is left. The peak position is consistent with the stellar position to within $2\arcsec$ (\emph{Herschel's} $1-\sigma$ pointing accuracy), so the observed position is consistent with the Hipparcos astrometry projected to December 2011. 
 The colour scale is in mJy/square arcsecond. The hatched circles show the average PACS beam FWHM of $6.7\arcsec$. The residuals after subtraction of the PSF are shown in the right panel. The contours are at 3, 6, 12, 24, 48, 100 in units of the pixel to pixel variation, given by $2.5 \times 10^{-5}$ or $2.5 \times 10^{-5}$mJy/arcsec$^2$. These residuals clearly show the detection of extended emission over and above that of the star.  }
\label{fig:kappacrb}
\end{figure*}

\begin{figure*}
\includegraphics[width=0.48\textwidth]{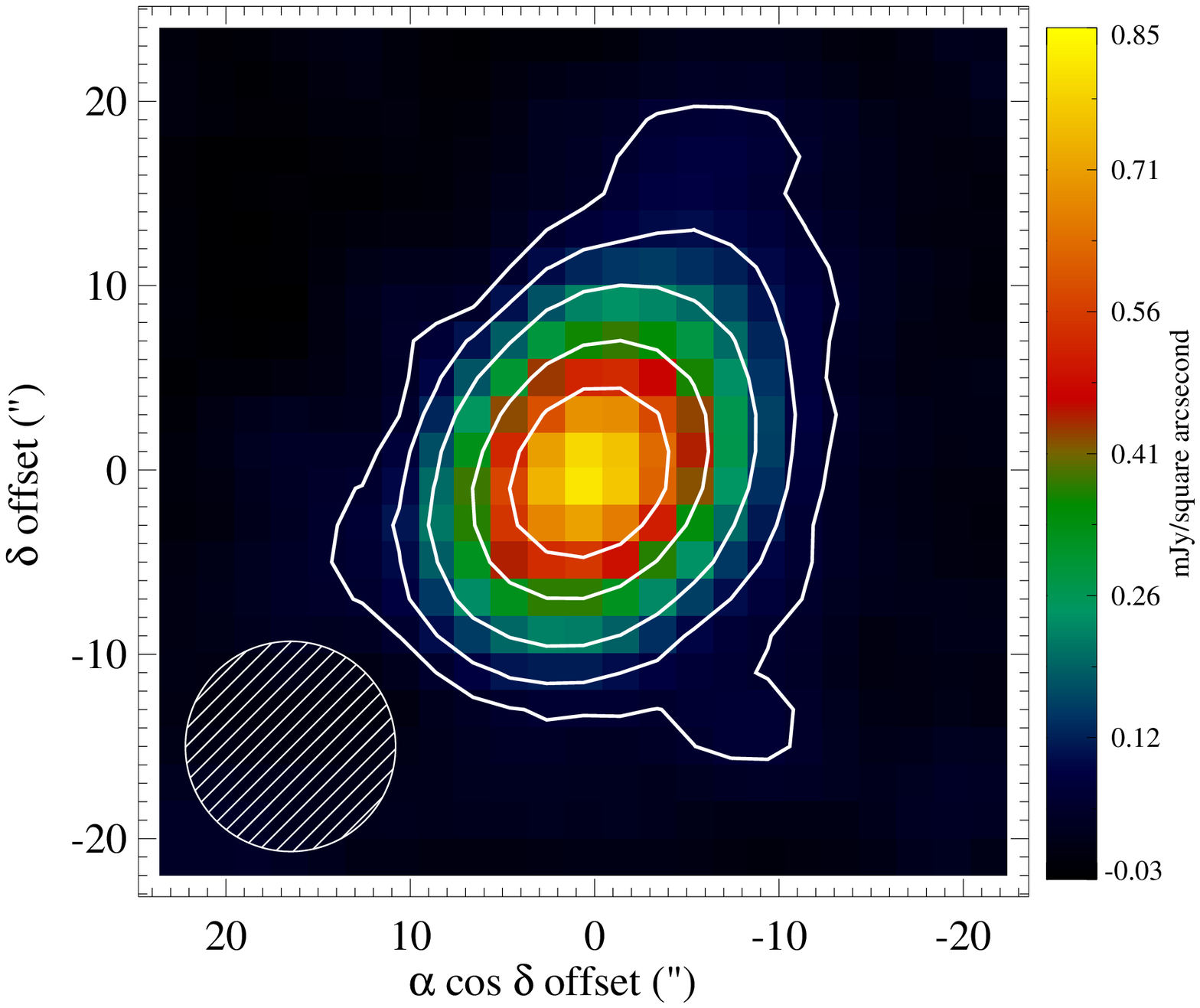}
\includegraphics[width=0.48\textwidth]{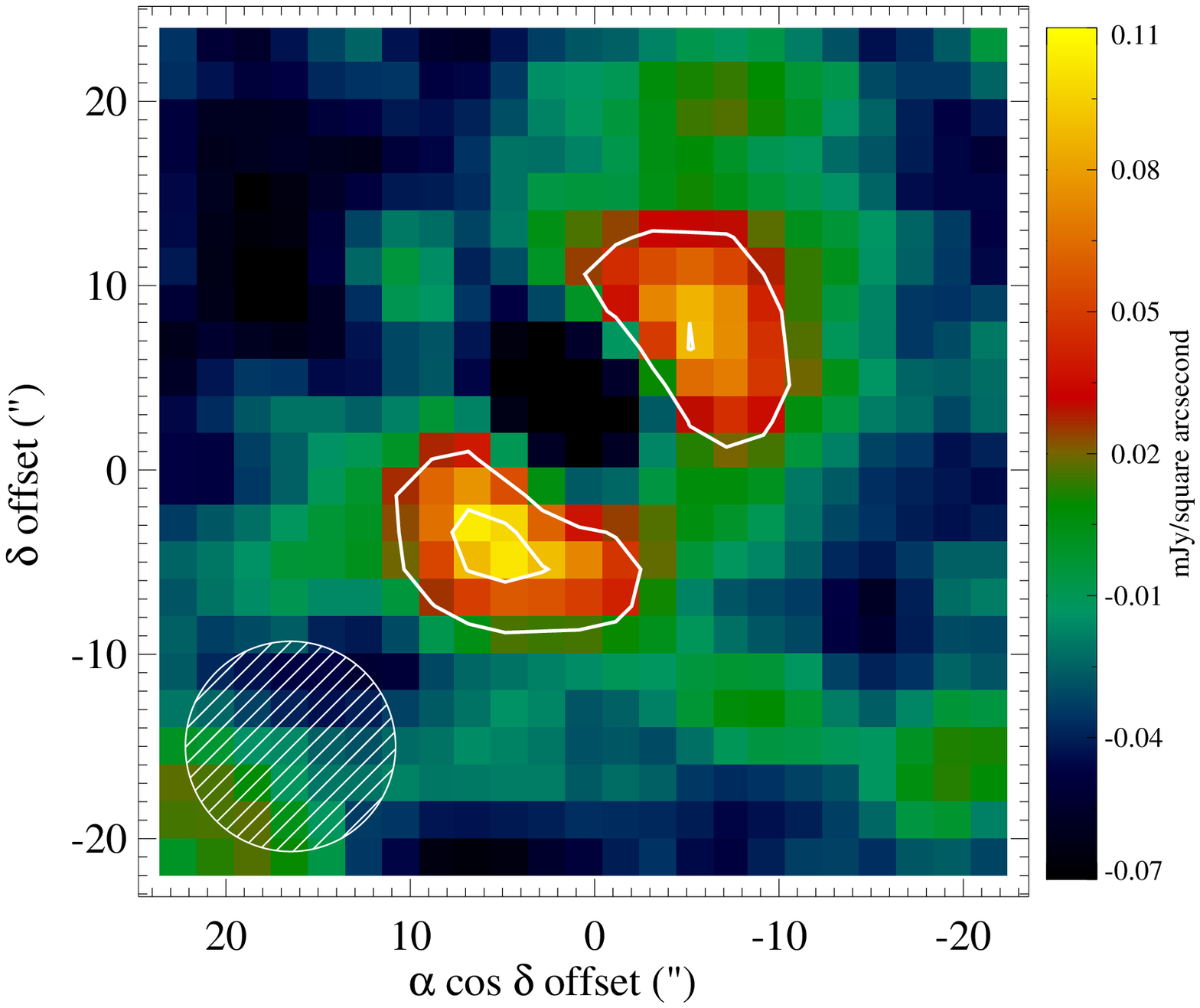}

\caption{ The same as Fig.~\ref{fig:kappacrb} except at 160$\mu$m. The average PACS beam FWHM is $11.4\arcsec$ and the contours 3, 6, 12, 24, 36 in units of the pixel to pixel sigma, given by $6.4 \times 10^{-5}$ or $2.6\times 10^{-4}$mJy/arcsec$^2$.  }
\label{fig:kappacrb160}
\end{figure*}


\section{Results}
\label{sec:results}
\subsection{Photometry and basic analysis of the \Herschel data}
\label{sec:basic}

By integrating the emission in an area surrounding the central source and comparing this to predictions for the stellar photosphere, calculated in the manner discussed in \S\ref{sec:sed}, we are able to calculate the level of excess flux. These values are quoted in Table~\ref{tab:flux}, where the disc is detected at $\sim17\sigma$ and $18 \sigma$ at 100 and $160\mu$m, respectively.

\Herschel PACS images at 100 and 160$\mu$m are shown in Figs.~\ref{fig:kappacrb} and ~\ref{fig:kappacrb160}. Firstly the observations are compared to those expected from a single point source, \ie unresolved emission. The residuals after subtracting a point source scaled to roughly the peak emission level are shown in the right hand panels of Figs.~\ref{fig:kappacrb} and ~\ref{fig:kappacrb160}. The residuals clearly show that the emission is extended. To learn further about the extended emission, we fitted a 2D Gaussian to both sources. We find emission that is elongated along a position angle of $\sim 145^\circ$, where the position angle is measured between North and the long axis of the ellipse, and East is positive.  The peak position is consistent with the stellar position to within $2''$ ($1$-$\sigma$ pointing accuracy). The full width half maximum of the emission is $11.2'' \pm 0.01'' \times 8.26'' \pm 0.01''$ at $100\mu$m and $14.9'' \pm 0.013'' \times 11.3'' \pm 0.017''$ at $160\mu$m. Given that the PSF is extended by $6.78'' \times 6.95''$ at $100\mu$m and $12.1''\times 10.7''$ at 160$\mu$m \citep{Kennedy_binary}\footnote{These sizes are slightly larger than quoted in the PACS Observer's Manual, because our data cannot be recentered on a scan by scan basis (i.e. \kcrb is much fainter than the PACS calibration targets used for PSF characterisation). The PSF sizes are quoted as minor x major axis size (i.e. reversed relative to the disk sizes) because the PSF tends to be elongated in the in-scan direction \citep{Kennedy_binary} and the $\kappa$ CrB disc is roughly perpendicular to the scan direction.}, the emission is clearly resolved at both wavelengths. The elliptical shape of such emission is difficult to reconcile with a spherically symmetrical shell, but could result from an azimuthally symmetrical disc, viewed along a line of sight inclined to the disc plane. In which case, the major to minor axis ratio implies that the disc would be inclined at an angle of $58^\circ \pm 1^\circ$ (100$\mu$m) or $48^\circ\pm 1^\circ$ ($160\mu$m), from face-on, where the uncertainties come straight from fitting the 2D Gaussian. The disc would have a deconvolved diameter of $\sim 280$ AU at $100\mu$m and 320AU at $160\mu$m.

\subsection{Detection limits on (planetary) companions}
\label{sec:detlimits}

The constant acceleration found by the radial velocity monitoring of \kcrbcomma can be used to place limits on the orbital parameters of the second companion to \kcrbdot The simplest assumption to make is that the planet is on a circular orbit, in which case a minimum limit on the companion's mass can be calculated by assuming that its gravity is responsible for the observed acceleration. This limit depends on the separation of the companion from the star, or its semi-major axis and is given by: 

\begin{equation}
m\sin i > 1.22 M_J \left(\frac{a}{12}\right)^2,
\label{eq:mmin}
\end{equation}
where the semi-major axis is $a$ in AU. 

\begin{figure}
\includegraphics[width=0.48\textwidth]{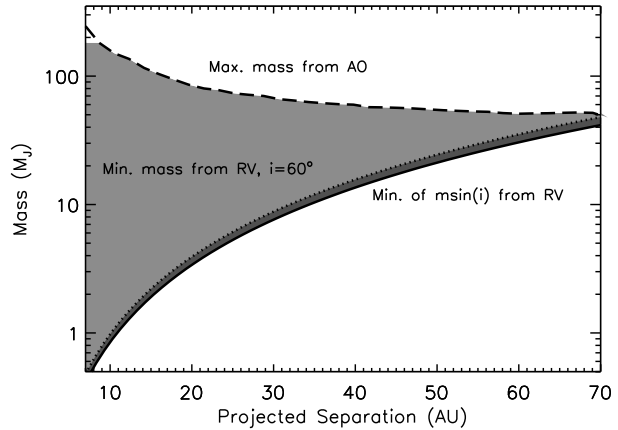}
\caption{The limits on the planet/companion mass and separation from the host star, based on the non-detections found in the AO imaging (black dashed line, see \S\ref{sec:AO}) and the radial velocity trends found in the Keck RV data (solid line- see \S\ref{sec:detlimits}). For the latter, the mass limit is the minimum mass, $m\sin i$ and the projected separations, the semi-major axis of the planet, assuming that it is on a circular orbit. In order to explain both sets of observations, \kcrb must have a companion that lies within the shaded area. The dotted line shows the minimum planet mass, based on the assumption that the planet's orbit is inclined by $60^\circ$ from face-on, in the same manner as the disc. }
\label{fig:det_lim}
\end{figure}

This limit is plotted in Fig.~\ref{fig:det_lim}. Our non-detection of the companion using high-contrast imaging places an upper limit on the possible mass of the companion, as a function of its projected separation from the star. Our ($5 \sigma$) sensitivity to off-axis sources, as a function of angular separation, calculated using the \cite{Baraffe2003} evolutionary models, is over-plotted in Fig.~\ref{fig:det_lim}. This was calculated using \kcrbs measured parallax ($\pi=32.79\pm 0.21$) and estimated isochronal age ($2.24\pm0.15$ Gyr). The largest uncertainty here is in the age of the primary star, followed by systematic errors in the thermal evolutionary models.

These non-detection limits complement the radial velocity measurements and place tight constraints on the properties of \kcrb c. The allowable mass and semi-major axis parameter space is shown by the shaded area in Fig.~\ref{fig:det_lim}. For instance, Doppler observations rule out the existence of extrasolar planets ($m<13M_J$) beyond $\sim40$ AU. The minimum possible mass of \kcrb c can be determined by considering that the minimum possible orbital period of the companion, even with an eccentricity of $e=1$, corresponds to the time baseline of the observations. In this case, the minimum period is therefore 8.09 years, which corresponds to a minimum semi-major axis of 7.3AU and thus, mass of $m \sin i > 0.5 M_J$.


\begin{figure}
\vspace{-100pt}
\includegraphics[width=0.45\textwidth]{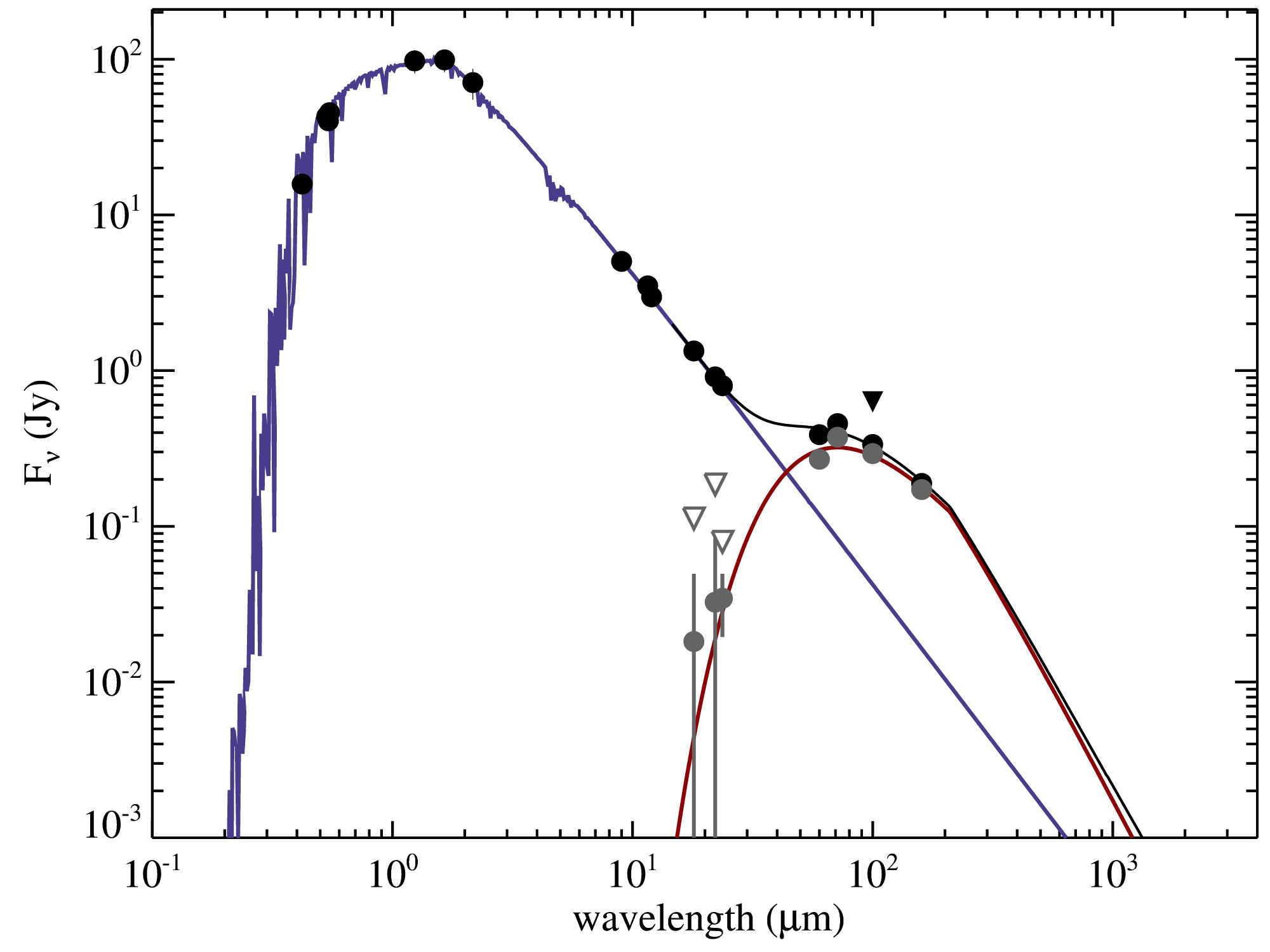}
\caption{SED for $\kappa$ Cr B. Photometry is shown as black dots or black triangles for upper limits. Disk (i.e. photosphere-subtracted)
fluxes and upper limits are shown as grey dots and open triangles. The stellar spectrum is shown as a blue line
and the modified black-body disk model as a red line, with the total shown as a black line. In the modified black-body model $\lambda_0$ was arbitrarily set at $210\mu$m and $\beta$ at 1, following \citet{wyattreview}.}
\label{fig:sed}
\end{figure}


\section{Models of the \Herschel data}
\label{sec:models}

The \Herschel images of \kcrb resemble images of many other debris discs around main-sequence stars. The confined, disc-like nature of the source suggests that we are observing the collisional remnants of planetesimals in a Kuiper-like belt. Although \kcrb has evolved off the main-sequence, its properties have not changed significantly from that of a main-sequence star, in particular, there is no expectation that the rate of stellar mass loss has increased beyond the gentle stellar winds of main-sequence stars \footnote{ Significant stellar mass loss that could produce an infra-red excess emission only occurs towards the tip of the giant branch and on the asymptotic giant branches. Evidence against increased stellar mass loss rates for sub-giant stars comes from the lack of a need to include any stellar mass loss in evolutionary models that fit observations of globular clusters \citep{Iben1967}}. Thus, although previous explanations for observations of giant stars with infrared excess have included sporadic mass ejections and interstellar clouds \citep{kimzuckermansilverston01, zuckermankim95, Jura99}, the evidence in the case of \kcrb is strongly in favour of emission from a debris disc. The only other possible source of emission at such wavelengths could be the companion(s), which the level and morphology of the excess emission lead us to believe is clearly not the origin in this case.


In this section we consider the \Herschel observations in their own right, comparing the observations with the emission from a model debris disc. The intention of this modelling is to derive as much information as possible regarding the disc structure from the \Herschel images. Given the limitations on the information available from these images, it will only really be possible to determine a rough size for the disc and place some broad constraints on its orientation. In \S\ref{sec:discussion} we discuss these models in the light of the limits on planetary companions in this system.

In order to model the emission from a debris disc, firstly, the contribution of the stellar emission must be accounted for. Optical and near-infrared photometry is collected from numerous catalogues \citep{Morel1978, Moshir1993, Hauck1997, Perryman1997, Hog2000, Cutri2003, Mermilliod1987, Ishihara2010}. These data were used to find the best fitting stellar model, using the PHOENIX Gaia grid \citep{Brott2005}, via a $\chi^2$ minimisation, as in \cite{Kennedy99Her, Kennedy_binary, Wyatt61Vir}. This method uses synthetic photometry over known bandpasses and has been validated against high S/N MIPS 24$\mu$m data for DEBRIS \citep{Matthews2010} targets, showing that the photospheric fluxes are accurate to a few percent for main-sequence, AFG-type, stars.


\subsection{Spectral Energy Distribution (SED)}
\label{sec:sed}

 The synthetic stellar spectrum is plotted in Fig.~\ref{fig:sed}. We have added to this the \Herschel PACS data, as well as archival Spitzer data \citep{Kalas_proposal} and data points from the IRAS faint source catalogue.  
There is clear evidence for excess emission above the predicted stellar spectrum, as can also be seen in Table.~\ref{tab:flux}.

The only information that can be obtained from just the spectral energy distribution (SED) is an estimate of the disc temperature. In order to determine this, we make the simplest possible assumption; that the dust grains emit like black-bodies. We anticipate that the inefficient emission properties of real grains reduce the flux at wavelengths longer than $\lambda_0$ by a factor $\left (\frac{\lambda}{\lambda_0}\right)^{-\beta}$, where we have introduced the free parameters $\lambda_0$ and $\beta$ to take this into account. Since, here we have detections at only four different wavelengths, $\lambda_0$ and $\beta$ are very poorly constrained, but nonetheless illustrative of the reduced emission anticipated at long wavelengths, that could be relevant for future observations, for example with ALMA.

The disc temperature can be determined from Fig.~\ref{fig:sed} using our modified black-body description. We find a temperature of $60 \pm 10$K, although $\beta$ and $\lambda_0$ remain unconstrained and it is clear that some discrepancy exists between the IRAS $60\mu$m and Spitzer MIPS $70\mu$m points. It is not possible to fit a modified black-body that agrees with both these points to within the uncertainties (which are smaller than the data points on the plot). We, therefore, assessed whether the Spitzer $70\mu$m point might be contaminated by background sources, but deem this to be unlikely as the flux varies by less than 4\%  between apertures of different sizes, whilst the uncertainty on each data point is 5\%. There is more reason to question the IRAS point, as there is a greater than 30\% variation in flux between the IRAS Point Source Catalogue (IRAS-PSC) and IRAS Faint Source Catalogue (IRAS-FSC), whilst the quoted uncertainties are 10\% in IRAS-PSC and 8\% in IRAS-FSC. In general, the IRAS-FSC is more reliable, and this is the point that we use, however, it may be that the uncertainty on this point should be increased above that quoted.

If this fit is correct and the dust indeed acts like a black body, a temperature of $60\pm 10$K would correspond to a radius of $75^{+35}_{-20}$AU. The large uncertainty in the disk temperature, and therefore black-body radius, arises because the temperature is degenerate with $\lambda_0$ and $\beta$. The inferred disc radius is therefore smaller than suggested by the images, which could arise due to the presence of small ($\mu$m) grains that are hotter than blackbodies. However, the temperature is sufficiently uncertain that this conclusion is not strong.

\begin{figure*}
\includegraphics[width=0.48\textwidth]{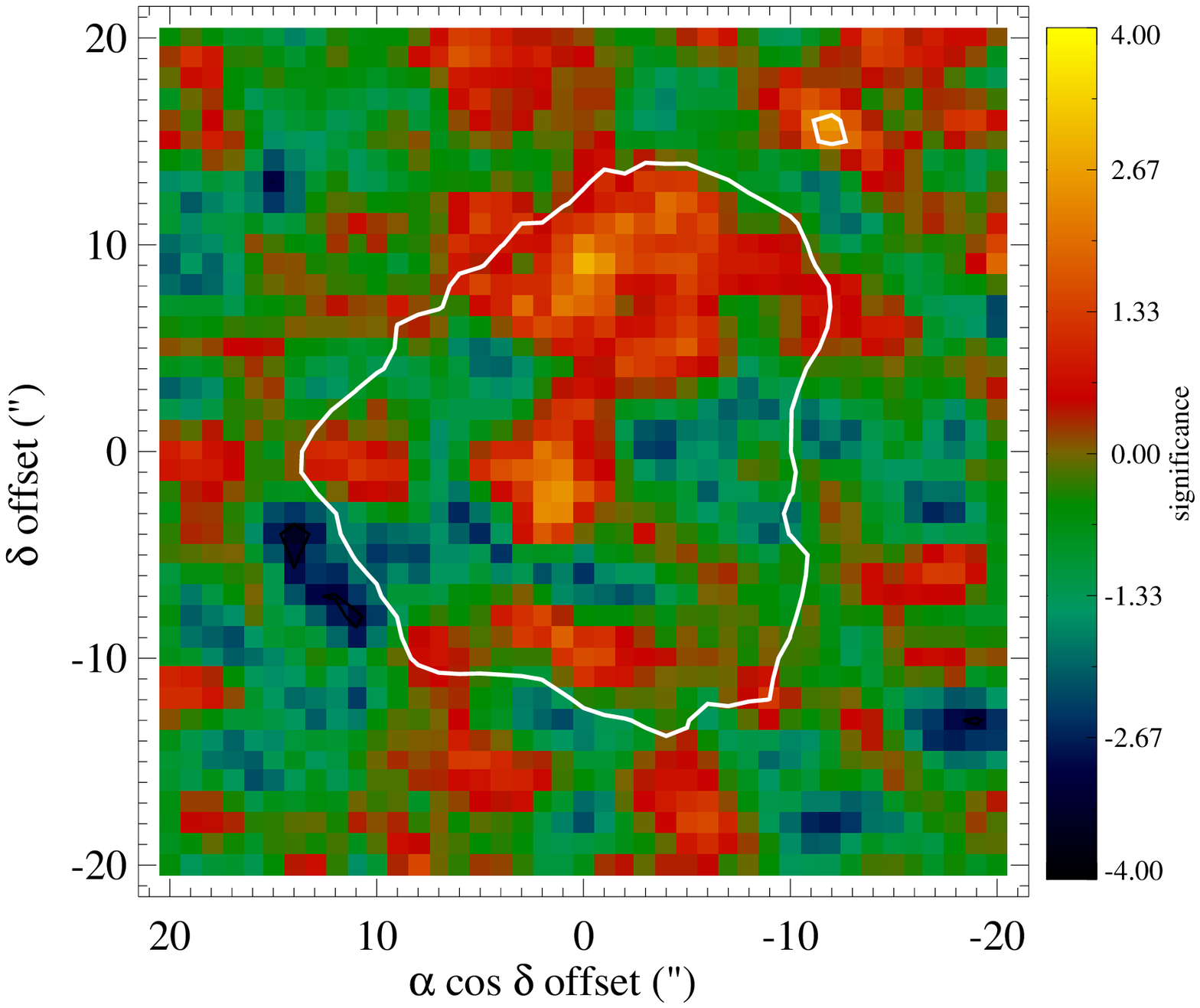}
\includegraphics[width=0.48\textwidth]{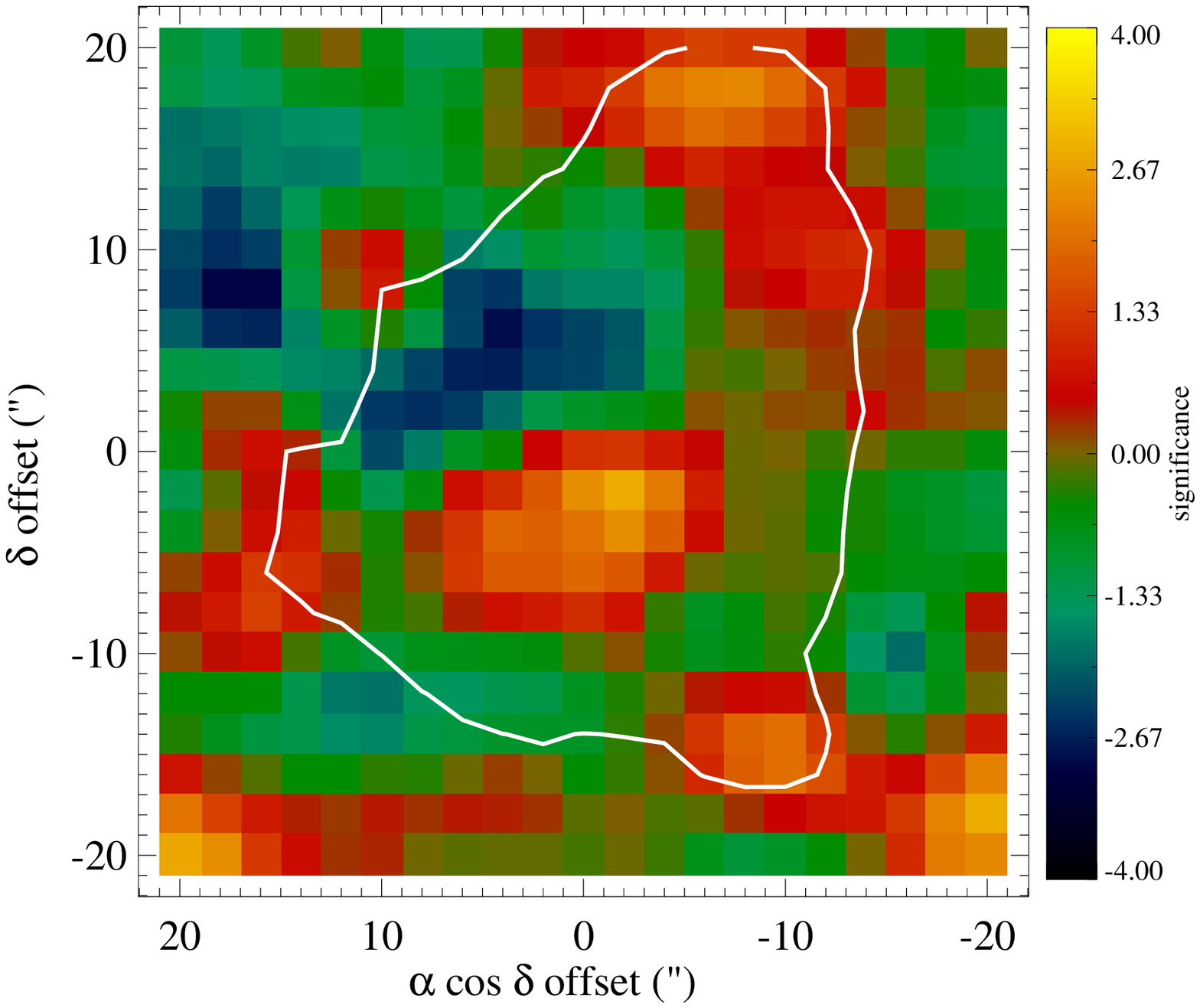}
\caption{The residuals, in units of significance of (image-model)/uncertainty, from the model fits with a single belt described in \S\ref{sec:modelkcrb} at 100 (left) and $160\mu$m (right). The models provide a good fit to the observations, the residuals are at low levels (note the scale), with the black contours showing  residuals of $3-\sigma$. The white contours show the $3-\sigma$ detection of the disc, as shown in Fig.~\ref{fig:kappacrb} and Fig.~\ref{fig:kappacrb160}. }
\label{fig:obs:kcrb}
\end{figure*}
\begin{figure*}
\includegraphics[width=0.48\textwidth]{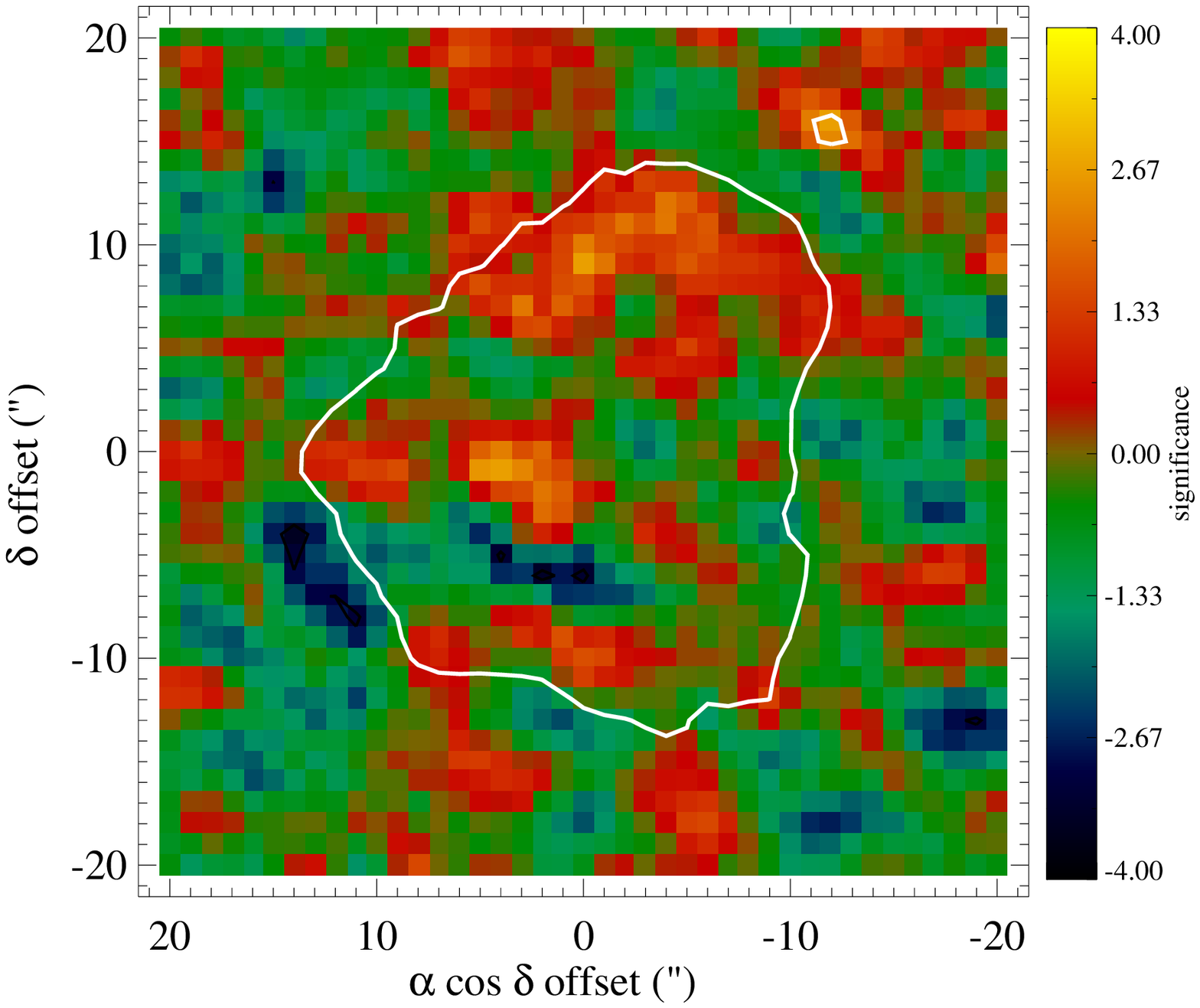}
\includegraphics[width=0.48\textwidth]{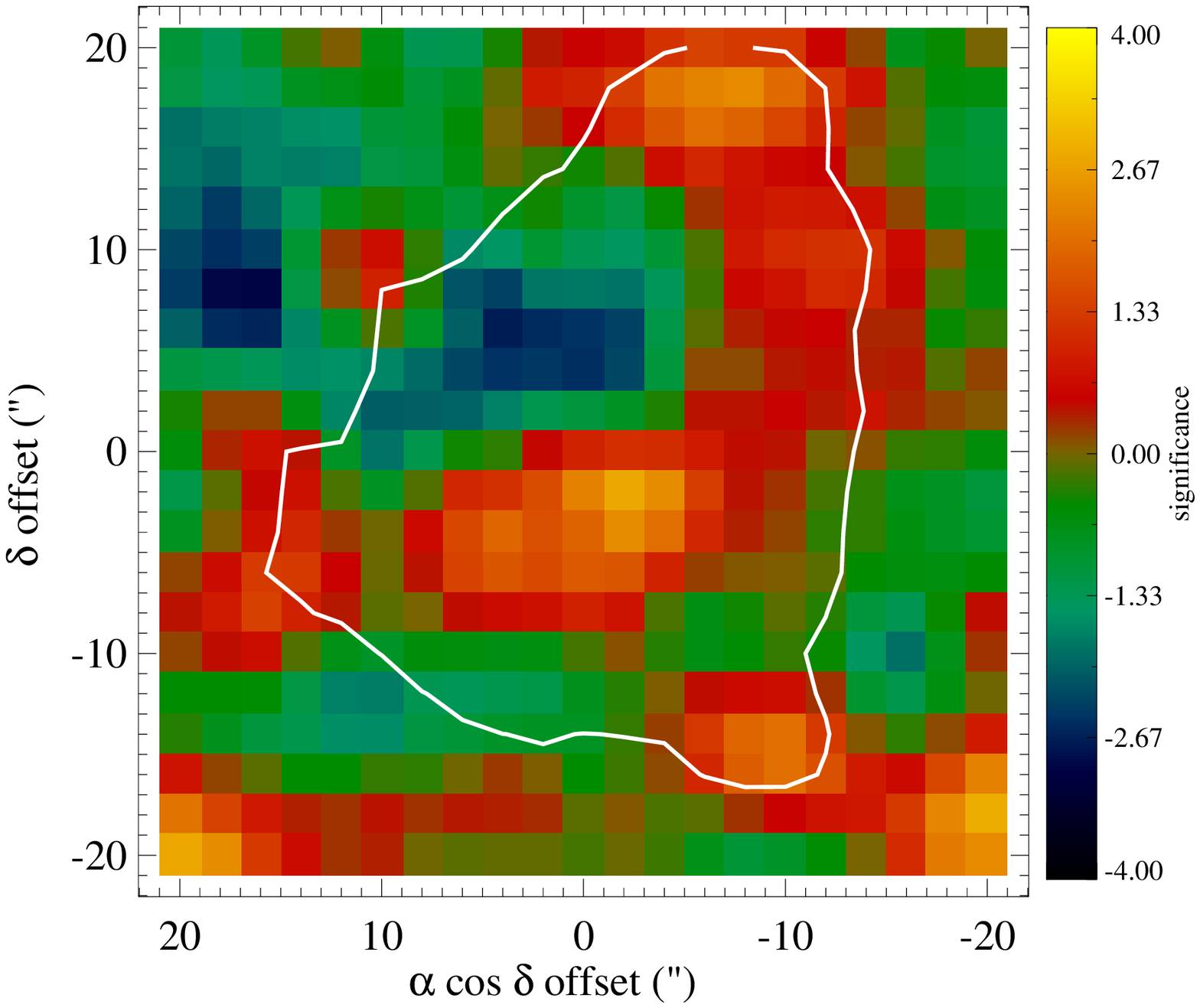}

\caption{The same as Fig.~\ref{fig:obs:kcrb}, but this time showing the residuals from the two belt fit.}

\label{fig:obs:2kcrb}
\end{figure*}


\subsection{A model of the \Herschel images} 
\label{sec:modelkcrb}
\kcrb is resolved in both of the \Herschel images. We, therefore, attempt to fit the observations  using a simple model. Firstly, the stellar photosphere is subtracted, such that we are left with emission from the debris disc, which can directly be compared with the emission from a model disc. 
The models we generate here use the methods described in \cite{Wyatt61Vir, Kennedy99Her, Kennedy_binary}. The basic idea is to determine the spatial distribution of the dust, which coupled with the grain emission properties at each wavelength can be used to produce a high resolution model image of the disc at each wavelength. In these models we make no attempts to constrain the grain properties or size distribution, since in the absence of spectral features this will be degenerate with other assumptions.  Instead we use the modified black-body prescription, outlined in \S\ref{sec:sed}. In this modified black-body prescription, the temperature of the disc is assumed to be $T=f_T T_{bb}$, where $T_{bb}= 278.3 \left (L_* / L_\odot \right)^{\frac{1}{4}} \left (r/AU \right )^{-\frac{1}{2}}= 521(r/AU)^{-\frac{1}{2}}$ is the temperature that a black-body at a distance r from the star of luminosity $L_*=12.3 L_\odot$ would achieve, and $f_T$ is a parameter of the model, related to the grain emission efficiency, the same as used in \cite{Wyatt61Vir} and \cite{Lestrade2012}.

The spatial distribution of the dust is assumed to be disk-like, with a small opening angle. It is characterised by the dust's face-on optical depth ($\tau$), which is parametrised as a function of radius by one or more power laws. The dust can be arranged in a single or multiple belts, characterised by their radial location, width, position angle and inclination to the line of sight. The quality of each model is evaluated using a sum of squared image-model differences approach as described in \cite{Wyatt61Vir}.

These models are defined by a large number of parameters, which means that the best fit model is not determined by undertaking a grid search of all possible parameter spaces, rather by a combination of by-eye fitting and least-squares minimisation. While we do not claim that the models are unique, we do show that they are plausible layouts of the debris disk structure. As we show below, we can reproduce the observed disk structure with several different models, and therefore consider the uncertainties on individual model-specific parameters relatively unimportant.

Given that our modelling approach does not include detailed grain properties, it is likely that the true disk emission spectrum is more complex than our simple modified black-body, for example including relatively narrow spectral features. Differences may also exist between different sets of photometry due to calibration offsets. As can be seen in the SED (Fig.~\ref{fig:sed}) the spectrum suggested by the four far-IR points is not well represented by the modified black-body, thus, it may be suffering from these problems. While these differences do not pose a major problem for SED modelling, they complicate image modelling since a small absolute offset caused by an attempt to achieve the best fit for all data can make an otherwise satisfactory image model to appear very poor when compared with the observed image. Our approach, therefore, follows \cite{Lohne2012}, who introduce small offsets at some wavelengths, so that the image models focus on fitting the radial distribution of the emission, rather than the overall disc flux, which can be biased due to the preceding factors. In this manner we could apply small modifications to each of the four far-IR points, however, for simplicity we found that it was only necessary to apply such a modification to the PACS $100\mu$m point, using a factor $C_{100}$.

Considering firstly a single dusty ring, our best fit model is a wide belt extending from 20 to 220AU, with an optical depth of $\tau = 2.5\times10^{-5} r^{0.5}$, a temperature profile of $T= 597 r^{-0.5}$ \ie $f_{T}=1.1$, $\lambda_0=70\mu$m, $\beta=0.6$ and $C_{100}=0.95$. These parameters are not particularly well constrained, for example, from the SED fitting we already concluded that $\lambda_0$ and $\beta$ are unconstrained, and reasonable fits to the data can be made for different values of the disc inner radius. However, a clear conclusion that a single, narrow belt does not fit this data is made. Better constrained are the inclination and position angle, for which we determine values of $59^\circ$ from face-on and $142^\circ$, respectively. The values quoted here are similar to those found by fitting an ellipse to the emission in \S\ref{sec:basic} and we consider these to be constrained to within $\pm10^\circ$. A reduced $\chi^2$ of 0.8 is calculated for this model, in the manner described in \cite{Wyatt61Vir}. \footnote{Although the number of degrees of freedom (7 weighted values) used to calculate this value may be misleading (see \cite{Wyatt61Vir}), a value close to 1 suggests a satisfactory fit.}
The residuals once this model fit is subtracted from the observations (Fig.~\ref{fig:obs:kcrb}) show that this model is a good fit to the data. The positive increase in optical depth with radius is unusual for debris disc models, however, we were unable to find a negative slope that provided a good fit to the data. This positive increase in surface density could be a real feature, related to the dynamics or stirring mechanism of the disc. Alternatively, it could provide an indication that this structure is an incorrect interpretation of the real disc structure.

Given that such a wide single belt fits the data, the possibility that the system could contain multiple belts is worthy of investigation. We therefore relax the constraints and allow the model to include two dust belts. So that there is no increase in the number of parameters required in order to obtain a fit, we fix various parameters. We assume that both belts are 10AU wide, that the temperature dependence is $T=521r^{-0.5}$ ($f_{T}=1$), $\lambda_0=70\mu$m and that each ring has constant optical depth with radius. We find $C_{100}=0.91$. The best fit model has two belts, centred on 41 and 165AU, with optical depths of $\tau= 2.7\times 10^{-4}$ and $\tau= 1\times 10^{-3}$ and $\beta=0.7$ and $1.0$, respectively. The disc orientation, with a position angle of $145^\circ$ and inclination of $60^\circ$ from face-on, remains close to the original estimation. The residuals for this model are shown in Fig.~\ref{fig:obs:2kcrb}, and a value of $\chi^2=0.7$ indicates that this is a good fit.

Both of these models provide equally good fits to the data. They both have a common radial scale for the dusty material (from around 20AU to around 200AU) and an inner hole. The differences between these models reflects our lack of knowledge of the distribution of the emission within this region. In fact, the inner hole need not be completely empty and the observations are equally consistent with a stirring model in which the disc extends from near to the inner planet out to large radial distances, but is collisionally eroded from inside-out. In such a model the majority of the emission is still from the 20-200AU region.

Planetesimals can be stirred as a result of the manner in which the protoplanetary gas disk was dispersed, the formation of Pluto-sized objects \citep{kenyonbromleyselfstirring}, or perturbations by a planetary companion \citep{alex}. Given the known orbit for the inner planetary companion, we can estimate the radial distance out to which the planet could have stirred the disc. Using Eq.15 of \cite{alex}, equating the stirring timescale, with the 2.5Gyr age of \kcrbcomma the 2.1$M_J$ planet could stir planetesimals out to around 70AU. The second, more distant companion might inevitably stir planetesimals beyond this distance, depending on the properties of the orbit.  
In order to fit such a model to the observations, we consider that the surface density distribution of dust is split into two regions, an inner region where the density increases with radial distance to the star and an outer region in which it decreases. We then determine the radius of this change in surface density, which for these images occurs at 81AU, with the outer region extending to 290AU. Our best fit finds $\tau=5.8\times 10^{-8} r^{2.3}$ inside of 81AU and $\tau=6.6\times 10^{-5} r^{-1}$ in the outer regions, where the slope of the optical depth in the outer region was fixed at $r^{-1}$, in both regions was fixed, with $f_T=1.15$. This model provides as good a fit to the data as the previous two models, as indicated by a $\chi^2$ value of $0.8$. Again, however, these parameters are not very well constrained. This model fits well with the planet stirring of planetesimals out to around 70AU by the known companion, but if our radial constraint is correct, the second companion must have orbital parameters close to the minimum possible ($a_{pl}=7.3$AU and $M_{pl}=0.5M_J$) and orbit on a very low eccentricity orbit ($e_{pl}<0.06$) in order that planet-stirring of planetesimals does not occur beyond $\sim80$AU.

\begin{figure*}

\includegraphics[width=0.89\textwidth]{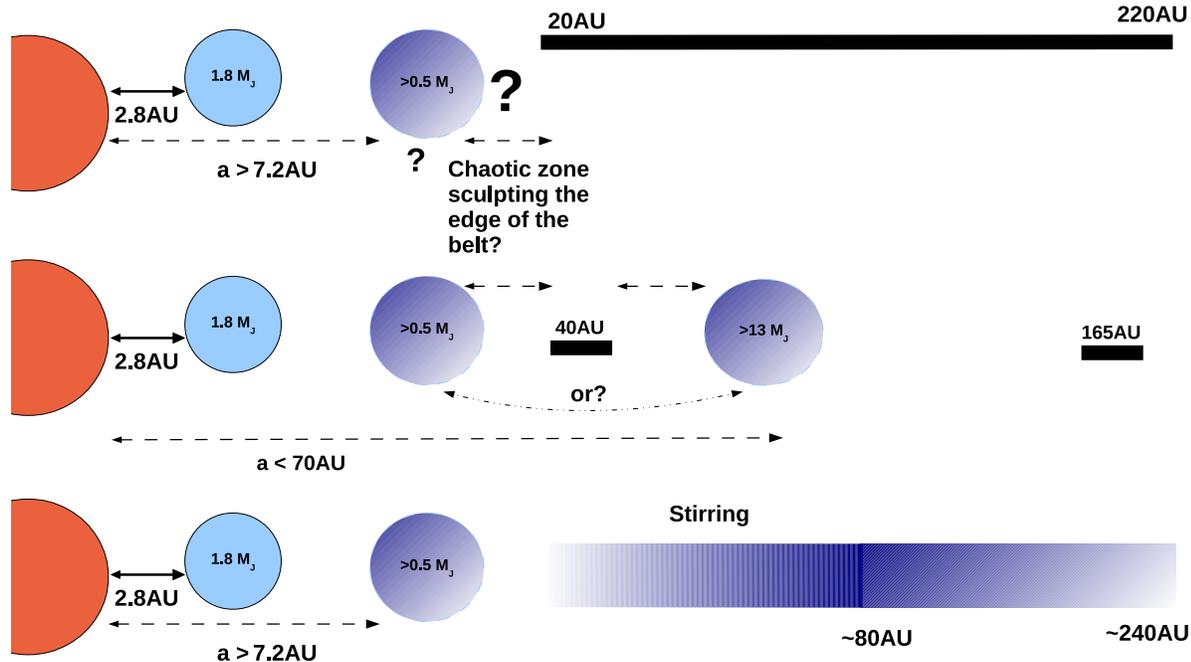}

\caption{A cartoon to illustrate the possible configurations of the \kcrb planetary system. The top panel illustrates the possibility that the second companion lies interior to the observed dusty material, that lies either in a single wide belt or is split into two narrow belts, as described by the best fit models in \S\ref{sec:modelkcrb}. The middle panel illustrates the possibility that the outer companion is in fact a binary and orbits between the two narrow dust belts. In the latter scenario the outer dusty belt would be a circumbinary debris disc. The lower panel illustrates the stirring model, in which the rate of dust production is maximum at $\sim80$AU. Diagram is not to scale. }

\label{fig:di_pl}
\end{figure*}

To summarise, we have presented three plausible models for the dusty material, as illustrated in Fig.~\ref{fig:di_pl}. All three models reproduce the \Herschel images and the SED. Without further observations we have no means of distinguishing between these models, nor ruling out alternative models.

\section{Discussion}
\label{sec:discussion}

In this work we have presented new \Herschel resolved images of the debris disc around \kcrbcomma alongside evidence for a second companion from Keck radial velocity data and upper limits on its mass from Keck AO imaging. Coupled with the known planetary companion from \cite{Johnson2008}, this allows us to constrain the structure of the \kcrb planetary system. Although our knowledge of the \kcrb planetary system has grown significantly from these observations, it is critically important to distinguish between the information that is well constrained from these observations and the more tentative conclusions that can be made. This will be discussed in the following section.

By constraining the structure of the \kcrb planetary system, we have provided an example of a planetary system around an intermediate mass subgiant, or `retired' A star, which in turn aids our understanding of the population of planetary systems around higher mass stars. In the second half of this discussion, we consider the impact of this study on our understanding of planetary systems in general.

\subsection{The Structure of the \kcrb Planetary System}
\label{sec:struct}

Firstly, we consider the \Herschel resolved images. One of the clearest conclusions that can be made regards the inclination of the circumstellar disc. 
The ellipsoidal nature of the excess emission can be clearly seen in the star-subtracted images shown in Fig.~\ref{fig:kappacrb}. The ellipse-like nature of the source suggests an azimuthally symmetric, circumstellar disc, viewed inclined to the line of sight. The inclination of this dust belt is reasonably well constrained from the images, at around $60^\circ$ from face-on, with a position angle of $145^\circ$. This in turn has implications for the second companion. If the planet and the dust disc formed out of the same proto-planetary disc and there have been no further interactions, it seems reasonable to assume that the inclination determined for the disc is also the inclination for the planet. This would mean that the planet's mass is a factor of $\frac{1}{\sin 60^\circ}\sim 1.15$ higher than the minimum mass of $m \sin i = 2.1 M_J$, with similar implications for the mass of the unconstrained companion. The dotted line on Fig.~\ref{fig:det_lim} shows this increase in the planet's mass. It should, however, be noted that the disc and the known planet are sufficiently well separated that it is possible that post formation processes have altered their inclinations from coplanar. Although, during the the 2.5Gyr lifetime of \kcrb, the secular perturbations from the known $2.1M_J$ companion could have aligned the disc out to $\sim120$AU, as suggested in \citep[e.g.][]{Augereau1999,Kennedy_binary}\footnote{This is the distance at which particles would have precessed once about the planet's orbital plane.}. This means that if the disc and planet were misaligned early in their evolution, the disc might appear warped at around $\sim120$AU, and our observations do not have sufficient resolution to detect this warp. 

The radial location and structure of the debris disc is not as well constrained as its inclination. In fact the main conclusion that can be made from these \Herschel observations is that there is clearly a dense population of dusty bodies orbiting \kcrbdot We suggest three structures that could explain the observations, although, based on the current data, we are unable to rule out other possibilities. Our first model is a continuous dust belt, extending from from 20 to 220AU. The second is two distinct narrow dust belts, centered on 41 and 165AU. The third is a stirring model, in which the disc is collisionally eroded from inside outwards. In this case, the dust production peaks at around 80AU. Currently our only source of further information regarding the structure of the planetary system comes from a consideration of the effect of the second companion on the dusty material.

Our constraints on the orbit of the second companion, shown in Fig.~\ref{fig:det_lim}, are not tight. We can, however, examine how it could fit into the three scenarios proposed. These are illustrated in Fig.~\ref{fig:di_pl}. In all three cases, the second companion could orbit close to the inner companion, interior to the dusty material, be it in a single belt, two distinct belts, or distributed in a wide belt that is stirred by the planet(s). The companion could even be responsible for sculpting the inner edge of the inner belt. 

On the other hand, if the dusty material were to be split into two narrow belts, we consider the possibility that the second companion could orbit between the belts. In this case, the companion could be responsible for clearing the region of material. The outer belt at 165AU is sufficiently far from the second companion (that must lie within 70AU) that it is unlikely to be affected by it. The stability of a belt between 36 and 46AU, however, is likely to be strongly influenced by the second companion. We note here, that our stated values for the exact radial position of the inner belt are not strongly constrained and that reasonable fits to the data can be made using different radial distributions of dust within the same region. Assuming that the companion is on a circular orbit, we can use the overlap of mean-motion resonances, outlined by the chaotic zone ($\delta a_{chaos}= 1.3 (\frac{m_{pl}}{M_*})^{2/7}$) \citep{Wisdom1980} to estimate the size of the region in which particles would be unstable and thus, we anticipate to be clear of dust. The derivation for the size of this zone is only strictly valid for companion masses that are significantly less than the stellar mass, which is not the case for companions $>>10M_J$, nonetheless it remains true that higher mass planets must be further from the disc, if they are not to destabilise it. Using this limit, we find that if the companion were to be at the maximum radial distance from the star of 70AU, and therefore have a mass between $\sim42M_J$ and $\sim47M_J$, a belt of dust could only be stable interior to this companion, if its outer edge were to be interior to 34AU. The values predicted for the inner belt in our model rule out such a scenario, however, given the uncertain nature of our radial constraints on the dusty belts, the only conclusion that we can make is that if the second companion is to orbit between the two belts of dust, it must be on an orbit with semi-major axis close to 70AU. The limits on the mass of the second companion are such that in order for it to orbit between the two belts it must be sub-stellar, rather than planetary in nature\footnote{Using $M>13M_J$, the deuterium burning limit, to define the maximum mass of a planet}. In this case, \kcrb would have an outer circumbinary debris disc, orbiting a main-sequence and brown dwarf binary, in which the main-sequence star has its own debris disc.


To summarise, \kcrb is orbited by significant levels of dusty material and at least two companions, at least one of which is planetary. The inclination of the planetary system is reasonably constrained with an inclination of $60^\circ$ from face-on and a position angle of $145^\circ$. We present three plausible models for the distribution of dust, a single wide belt, two narrow belts or a stirring model. Further resolved imaging of this source in the infrared or sub-mm, combined with further radial velocity monitoring of \kcrb to constrain any outer planetary companions are required in order to be able to fully constrain the structure of this planetary system.

\subsection{\kcrb in context}
\label{sec:context}

The \kcrb planetary system is special, firstly, because it is a rare example of a system where both planets and planetesimal belts have been detected orbiting an intermediate mass star ($m>1.4M_\odot$) and, secondly, because it is a unique example of a debris disc around a subgiant. \kcrb is the only $>1.4M_\odot$ star with a giant planet inside of 8AU and a resolved image of a debris disc.

The evolution of the star should not, on the subgiant branch, have had an effect on the planetary system, particularly not the outer regions of the planetary region studied in this work. The main difference between this and most debris discs observed around main-sequence stars, is that the debris disc must have survived the entire main-sequence lifetime. \kcrb has an age of 2.5 Gyr \citep{Johnson08}, most of which was spent on the main-sequence. Models for the collisional evolution of debris discs show a decrease in their fractional luminosity with time, as collisions erode the material in the disc \citep{Wyatt07hot}. An extension of these models, that also include the evolution of the star \citep{bonsor10}, shows that this collisional evolution is the main factor that diminishes the detectability of debris discs around subgiants. A population survey for debris discs around subgiants is required to confirm whether \kcrb is unusual in retaining its debris disc until the subgiant branch. The fact that we detect a debris disc at this epoch implies that \kcrb did not suffer an event similar to the Late Heavy Bombardment that cleared our Solar System \citep{Booth09}.

 The main advantage of the subgiant nature of this star is that it enables radial velocity techniques to find planetary companions in a manner that would not have been possible whilst this star was on the main-sequence. This raises the question as to whether other main-sequence A stars that have debris discs similar to that of \kcrb may also have undetected planetary companions. For example, if \kcrb has a single wide belt, then it is not dissimilar to the single wide planetesimal belt of Vega \citep{Aumann1984, vegasu05,muller}, which raises the question of whether Vega potentially has inner planetary companions, that cannot currently be detected. If on the other hand, the dust in \kcrb is split into two narrow belts, similarities can be seen with HR 8799 \citep{Hr8799su,hr8799detection08,4thplanetHR8799} in that the dust belts are separated by planetary companion(s). In turn, this comparison could suggest that \kcrb may have further, undetected companions.


\section{Conclusions}
\label{sec:conclusions}
We have presented observational evidence for the structure of the \kcrb planetary system, a unique example of a debris disc around a subgiant and a rare example of an intermediate mass star, where both planets and planetesimal belts have been detected. \Herschel observations show high levels of excess emission, evidence generally taken for the presence of a debris disc. Following detailed modelling of the \Herschel observations we suggest three possible structures for the dusty material. Either a single wide belt, extending between $\sim20$ and $\sim 220$AU, two narrow belts, centered on $\sim40$ and $\sim165$AU, or a stirred belt, most probably stirred by the planetary companion(s), in which the dust production rates peak at $\sim80$AU. Our best constraint is on the inclination and position angle of the disc, which we place at $60^\circ$ from face-on and $145^\circ$, respectively, to within $\sim10^\circ$.

Alongside the \Herschel observations, we have presented evidence for the presence of a second companion to \kcrb found in the continued radial velocity monitoring of this star. This is in addition to the $m \sin i = 2.1M_J$ planet, at 2.8AU, detected by \cite{Johnson08}. An upper limit on the mass of this companion was found by its non-detection in AO imaging taken with Keck. The details are shown in Fig.~\ref{fig:det_lim}. In terms of our suggested models for the dusty material orbiting \kcrb, one possibility is that both companions lie interior to all the dusty material, potentially sculpting the inner edge of the inner belt. Alternatively, if the dusty material is split into two belts, we cannot rule out the possibility that the second companion could lie between these belts, giving \kcrb an intriguing structure with both a circumbinary and circumprimary debris disc.

As the first example of a planetary system orbiting a subgiant, a more detailed population study is required to determine whether or not \kcrb is unusual, nonetheless, this work suggests that \kcrb did not suffer any dynamical instability that cleared out its planetary system, similar to the Late Heavy Bombardment. As the first example of a $>1.4M_\odot$ star, with a giant planet interior to 8AU, where there is also resolved imaging of a debris disc, \kcrb provides a good example system from which to further our understanding of planetary systems around intermediate mass stars.

\section{Acknowledgements} 
We thank the referee for comments that impoved the quality of this manuscript. We thank  Steve Ertel and Jean-Charles Augereau for useful discussions that benefitted this work. 
AB acknowledges the support of the ANR-2010 BLAN-0505-01 (EXOZODI). 
MCW and GK are grateful for support from the European Union through ERC grant number 279973.

\bibliographystyle{mn}

\bibliography{ref}

\begin{thebibliography}{66}
\expandafter\ifx\csname natexlab\endcsname\relax\def\natexlab#1{#1}\fi

\bibitem[{{Augereau} {et~al.}(1999){Augereau}, {Lagrange}, {Mouillet},
  {Papaloizou}, \& {Grorod}}]{Augereau1999}
{Augereau} J.~C., {Lagrange} A.~M., {Mouillet} D., {Papaloizou} J.~C.~B.,
  {Grorod} P.~A., 1999, \aap, 348, 557

\bibitem[{{Augereau} {et~al.}(2001){Augereau}, {Nelson}, {Lagrange},
  {Papaloizou}, \& {Mouillet}}]{Augereau01}
{Augereau} J.~C., {Nelson} R.~P., {Lagrange} A.~M., {Papaloizou} J.~C.~B.,
  {Mouillet} D., 2001, \aap, 370, 447

\bibitem[{{Aumann} {et~al.}(1984){Aumann}, {Beichman}, {Gillett}, {de Jong},
  {Houck}, {Low}, {Neugebauer}, {Walker}, \& {Wesselius}}]{Aumann1984}
{Aumann} H.~H., {Beichman} C.~A., {Gillett} F.~C., {de Jong} T., {Houck} J.~R.,
  {Low} F.~J., {Neugebauer} G., {Walker} R.~G., {Wesselius} P.~R., 1984, \apjl,
  278, L23

\bibitem[{{Baraffe} {et~al.}(2003){Baraffe}, {Chabrier}, {Barman}, {Allard}, \&
  {Hauschildt}}]{Baraffe2003}
{Baraffe} I., {Chabrier} G., {Barman} T.~S., {Allard} F., {Hauschildt} P.~H.,
  2003, \aap, 402, 701

\bibitem[{{Bonsor} \& {Wyatt}(2010)}]{bonsor10}
{Bonsor} A., {Wyatt} M., 2010, \mnras, 409, 1631

\bibitem[{{Booth} {et~al.}(2012){Booth}, {Kennedy}, {Sibthorpe}, {Matthews},
  {Wyatt}, {Duch{\^e}ne}, {Kavelaars}, {Rodriguez}, {Greaves}, {Koning},
  {Vican}, {Rieke}, {Su}, {Moro-Mart{\'{\i}}n}, \& {Kalas}}]{Booth2012}
{Booth} M., {Kennedy} G., {Sibthorpe} B., {Matthews} B.~C., {Wyatt} M.~C.,
  {Duch{\^e}ne} G., {Kavelaars} J.~J., {Rodriguez} D., {Greaves} J.~S.,
  {Koning} A., {Vican} L., {Rieke} G.~H., {Su} K.~Y.~L., {Moro-Mart{\'{\i}}n}
  A., {Kalas} P., 2012, \mnras, 77

\bibitem[{{Booth} {et~al.}(2009){Booth}, {Wyatt}, {Morbidelli},
  {Moro-Mart{\'{\i}}n}, \& {Levison}}]{Booth09}
{Booth} M., {Wyatt} M.~C., {Morbidelli} A., {Moro-Mart{\'{\i}}n} A., {Levison}
  H.~F., 2009, \mnras, 399, 385

\bibitem[{{Bowler} {et~al.}(2010){Bowler}, {Johnson}, {Marcy}, {Henry}, {Peek},
  {Fischer}, {Clubb}, {Liu}, {Reffert}, {Schwab}, \& {Lowe}}]{Bowler2010}
{Bowler} B.~P., {Johnson} J.~A., {Marcy} G.~W., {Henry} G.~W., {Peek} K.~M.~G.,
  {Fischer} D.~A., {Clubb} K.~I., {Liu} M.~C., {Reffert} S., {Schwab} C.,
  {Lowe} T.~B., 2010, \apj, 709, 396

\bibitem[{{Brott} \& {Hauschildt}(2005)}]{Brott2005}
{Brott} I., {Hauschildt} P.~H., 2005, 576, 565

\bibitem[{{Chiang} {et~al.}(2009){Chiang}, {Kite}, {Kalas}, {Graham}, \&
  {Clampin}}]{chiang_fom}
{Chiang} E., {Kite} E., {Kalas} P., {Graham} J.~R., {Clampin} M., 2009, \apj,
  693, 734

\bibitem[{{Churcher} {et~al.}(2011){Churcher}, {Wyatt}, \&
  {Smith}}]{Churcher10}
{Churcher} L., {Wyatt} M., {Smith} R., 2011, \mnras, 410, 2

\bibitem[{{Crepp} {et~al.}(2012){Crepp}, {Johnson}, {Howard}, {Marcy},
  {Fischer}, {Hillenbrand}, {Yantek}, {Delaney}, {Wright}, {Isaacson}, \&
  {Montet}}]{Crepp2012}
{Crepp} J.~R., {Johnson} J.~A., {Howard} A.~W., {Marcy} G.~W., {Fischer} D.~A.,
  {Hillenbrand} L.~A., {Yantek} S.~M., {Delaney} C.~R., {Wright} J.~T.,
  {Isaacson} H.~T., {Montet} B.~T., 2012, ArXiv e-prints

\bibitem[{{Cutri} {et~al.}(2003){Cutri}, {Skrutskie}, {van Dyk}, {Beichman},
  {Carpenter}, {Chester}, {Cambresy}, {Evans}, {Fowler}, {Gizis}, {Howard},
  {Huchra}, {Jarrett}, {Kopan}, {Kirkpatrick}, {Light}, {Marsh}, {McCallon},
  {Schneider}, {Stiening}, {Sykes}, {Weinberg}, {Wheaton}, {Wheelock}, \&
  {Zacarias}}]{Cutri2003}
{Cutri} R.~M., {Skrutskie} M.~F., {van Dyk} S., {Beichman} C.~A., {Carpenter}
  J.~M., {Chester} T., {Cambresy} L., {Evans} T., {Fowler} J., {Gizis} J.,
  {Howard} E., {Huchra} J., {Jarrett} T., {Kopan} E.~L., {Kirkpatrick} J.~D.,
  {Light} R.~M., {Marsh} K.~A., {McCallon} H., {Schneider} S., {Stiening} R.,
  {Sykes} M., {Weinberg} M., {Wheaton} W.~A., {Wheelock} S., {Zacarias} N.,
  2003, VizieR Online Data Catalog, 2246, 0

\bibitem[{{Fruchter} \& {Hook}(2002)}]{Fruchter_Hook2002}
{Fruchter} A.~S., {Hook} R.~N., 2002, \pasp, 114, 144

\bibitem[{{Galland} {et~al.}(2005){Galland}, {Lagrange}, {Udry}, {Chelli},
  {Pepe}, {Queloz}, {Beuzit}, \& {Mayor}}]{Galland05}
{Galland} F., {Lagrange} A., {Udry} S., {Chelli} A., {Pepe} F., {Queloz} D.,
  {Beuzit} J., {Mayor} M., 2005, \aap, 443, 337

\bibitem[{{Girardi} {et~al.}(2002){Girardi}, {Bertelli}, {Bressan}, {Chiosi},
  {Groenewegen}, {Marigo}, {Salasnich}, \& {Weiss}}]{Girardi2002}
{Girardi} L., {Bertelli} G., {Bressan} A., {Chiosi} C., {Groenewegen} M.~A.~T.,
  {Marigo} P., {Salasnich} B., {Weiss} A., 2002, \aap, 391, 195

\bibitem[{{Hauck} \& {Mermilliod}(1997)}]{Hauck1997}
{Hauck} B., {Mermilliod} M., 1997, VizieR Online Data Catalog, 2215, 0

\bibitem[{{H{\o}g} {et~al.}(2000){H{\o}g}, {Fabricius}, {Makarov}, {Urban},
  {Corbin}, {Wycoff}, {Bastian}, {Schwekendiek}, \& {Wicenec}}]{Hog2000}
{H{\o}g} E., {Fabricius} C., {Makarov} V.~V., {Urban} S., {Corbin} T., {Wycoff}
  G., {Bastian} U., {Schwekendiek} P., {Wicenec} A., 2000, \aap, 355, L27

\bibitem[{{Iben}(1967)}]{Iben1967}
{Iben} Jr. I., 1967, \araa, 5, 571

\bibitem[{{Ishihara} {et~al.}(2010){Ishihara}, {Onaka}, {Kataza}, {Salama},
  {Alfageme}, {Cassatella}, {Cox}, {Garc{\'{\i}}a-Lario}, {Stephenson},
  {Cohen}, {Fujishiro}, {Fujiwara}, {Hasegawa}, {Ita}, {Kim}, {Matsuhara},
  {Murakami}, {M{\"u}ller}, {Nakagawa}, {Ohyama}, {Oyabu}, {Pyo}, {Sakon},
  {Shibai}, {Takita}, {Tanab{\'e}}, {Uemizu}, {Ueno}, {Usui}, {Wada},
  {Watarai}, {Yamamura}, \& {Yamauchi}}]{Ishihara2010}
{Ishihara} D., {Onaka} T., {Kataza} H., {Salama} A., {Alfageme} C.,
  {Cassatella} A., {Cox} N., {Garc{\'{\i}}a-Lario} P., {Stephenson} C., {Cohen}
  M., {Fujishiro} N., {Fujiwara} H., {Hasegawa} S., {Ita} Y., {Kim} W.,
  {Matsuhara} H., {Murakami} H., {M{\"u}ller} T.~G., {Nakagawa} T., {Ohyama}
  Y., {Oyabu} S., {Pyo} J., {Sakon} I., {Shibai} H., {Takita} S., {Tanab{\'e}}
  T., {Uemizu} K., {Ueno} M., {Usui} F., {Wada} T., {Watarai} H., {Yamamura}
  I., {Yamauchi} C., 2010, \aap, 514, A1

\bibitem[{{Johnson} {et~al.}(2010){Johnson}, {Aller}, {Howard}, \&
  {Crepp}}]{Johnson_planetpopulation}
{Johnson} J.~A., {Aller} K.~M., {Howard} A.~W., {Crepp} J.~R., 2010, \pasp,
  122, 905

\bibitem[{{Johnson} {et~al.}(2007){Johnson}, {Fischer}, {Marcy}, {Wright},
  {Driscoll}, {Butler}, {Hekker}, {Reffert}, \& {Vogt}}]{Johnson07}
{Johnson} J.~A., {Fischer} D.~A., {Marcy} G.~W., {Wright} J.~T., {Driscoll} P.,
  {Butler} R.~P., {Hekker} S., {Reffert} S., {Vogt} S.~S., 2007, \apj, 665, 785

\bibitem[{{Johnson} {et~al.}(2006){Johnson}, {Marcy}, {Fischer}, {Henry},
  {Wright}, {Isaacson}, \& {McCarthy}}]{Johnson06}
{Johnson} J.~A., {Marcy} G.~W., {Fischer} D.~A., {Henry} G.~W., {Wright} J.~T.,
  {Isaacson} H., {McCarthy} C., 2006, \apj, 652, 1724

\bibitem[{{Johnson} {et~al.}(2008{\natexlab{a}}){Johnson}, {Marcy}, {Fischer},
  {Wright}, {Reffert}, {Kregenow}, {Williams}, \& {Peek}}]{Johnson08}
{Johnson} J.~A., {Marcy} G.~W., {Fischer} D.~A., {Wright} J.~T., {Reffert} S.,
  {Kregenow} J.~M., {Williams} P.~K.~G., {Peek} K.~M.~G., 2008{\natexlab{a}},
  \apj, 675, 784

\bibitem[{{Johnson} {et~al.}(2008{\natexlab{b}}){Johnson}, {Marcy}, {Fischer},
  {Wright}, {Reffert}, {Kregenow}, {Williams}, \& {Peek}}]{Johnson2008}
---, 2008{\natexlab{b}}, \apj, 675, 784

\bibitem[{{Jura}(1999)}]{Jura99}
{Jura} M., 1999, \apj, 515, 706

\bibitem[{{Kalas} \& {Graham}(2008)}]{Kalas_proposal}
{Kalas} P., {Graham} J., 2008, Spitzer Proposal, 50737

\bibitem[{{Kalas} {et~al.}(2008){Kalas}, {Graham}, {Chiang}, {Fitzgerald},
  {Clampin}, {Kite}, {Stapelfeldt}, {Marois}, \& {Krist}}]{fomb2008}
{Kalas} P., {Graham} J.~R., {Chiang} E., {Fitzgerald} M.~P., {Clampin} M.,
  {Kite} E.~S., {Stapelfeldt} K., {Marois} C., {Krist} J., 2008, Science, 322,
  1345

\bibitem[{{Kennedy} \& {Kenyon}(2008)}]{Kennedy_Kenyon2008}
{Kennedy} G.~M., {Kenyon} S.~J., 2008, \apj, 673, 502

\bibitem[{{Kennedy} {et~al.}(2012{\natexlab{a}}){Kennedy}, {Wyatt},
  {Sibthorpe}, {Duch{\^e}ne}, {Kalas}, {Matthews}, {Greaves}, {Su}, \&
  {Fitzgerald}}]{Kennedy99Her}
{Kennedy} G.~M., {Wyatt} M.~C., {Sibthorpe} B., {Duch{\^e}ne} G., {Kalas} P.,
  {Matthews} B.~C., {Greaves} J.~S., {Su} K.~Y.~L., {Fitzgerald} M.~P.,
  2012{\natexlab{a}}, \mnras, 421, 2264

\bibitem[{{Kennedy} {et~al.}(2012{\natexlab{b}}){Kennedy}, {Wyatt},
  {Sibthorpe}, {Phillips}, {Matthews}, \& {Greaves}}]{Kennedy_binary}
{Kennedy} G.~M., {Wyatt} M.~C., {Sibthorpe} B., {Phillips} N.~M., {Matthews}
  B.~C., {Greaves} J.~S., 2012{\natexlab{b}}, \mnras, 426, 2115

\bibitem[{{Kenyon} \& {Bromley}(2004)}]{kenyonbromleyselfstirring}
{Kenyon} S.~J., {Bromley} B.~C., 2004, \aj, 127, 513

\bibitem[{{Kim} {et~al.}(2001){Kim}, {Zuckerman}, \&
  {Silverstone}}]{kimzuckermansilverston01}
{Kim} S.~S., {Zuckerman} B., {Silverstone} M., 2001, \apj, 550, 1000

\bibitem[{{Lafreni{\`e}re} {et~al.}(2007){Lafreni{\`e}re}, {Marois}, {Doyon},
  {Nadeau}, \& {Artigau}}]{Lafreniere2007}
{Lafreni{\`e}re} D., {Marois} C., {Doyon} R., {Nadeau} D., {Artigau} {\'E}.,
  2007, \apj, 660, 770

\bibitem[{{Lagrange} {et~al.}(2010){Lagrange}, {Bonnefoy}, {Chauvin}, {Apai},
  {Ehrenreich}, {Boccaletti}, {Gratadour}, {Rouan}, {Mouillet}, {Lacour}, \&
  {Kasper}}]{Lagrange2010}
{Lagrange} A., {Bonnefoy} M., {Chauvin} G., {Apai} D., {Ehrenreich} D.,
  {Boccaletti} A., {Gratadour} D., {Rouan} D., {Mouillet} D., {Lacour} S.,
  {Kasper} M., 2010, Science, 329, 57

\bibitem[{{Lagrange} {et~al.}(2009){Lagrange}, {Desort}, {Galland}, {Udry}, \&
  {Mayor}}]{Lagrange09}
{Lagrange} A., {Desort} M., {Galland} F., {Udry} S., {Mayor} M., 2009, \aap,
  495, 335

\bibitem[{{Lagrange} {et~al.}(2012){Lagrange}, {Milli}, {Boccaletti}, {Lacour},
  {Thebault}, {Chauvin}, {Mouillet}, {Augereau}, {Bonnefoy}, {Ehrenreich}, \&
  {Kral}}]{Lagrange2012}
{Lagrange} A.-M., {Milli} J., {Boccaletti} A., {Lacour} S., {Thebault} P.,
  {Chauvin} G., {Mouillet} D., {Augereau} J.~C., {Bonnefoy} M., {Ehrenreich}
  D., {Kral} Q., 2012, \aap, 546, A38

\bibitem[{{Lestrade} {et~al.}(2012){Lestrade}, {Matthews}, {Sibthorpe},
  {Kennedy}, {Wyatt}, {Bryden}, {Greaves}, {Thilliez}, {Moro-Mart{\'{\i}}n},
  {Booth}, {Dent}, {Duch{\^e}ne}, {Harvey}, {Horner}, {Kalas}, {Kavelaars},
  {Phillips}, {Rodriguez}, {Su}, \& {Wilner}}]{Lestrade2012}
{Lestrade} J.-F., {Matthews} B.~C., {Sibthorpe} B., {Kennedy} G.~M., {Wyatt}
  M.~C., {Bryden} G., {Greaves} J.~S., {Thilliez} E., {Moro-Mart{\'{\i}}n} A.,
  {Booth} M., {Dent} W.~R.~F., {Duch{\^e}ne} G., {Harvey} P.~M., {Horner} J.,
  {Kalas} P., {Kavelaars} J.~J., {Phillips} N.~M., {Rodriguez} D.~R., {Su}
  K.~Y.~L., {Wilner} D.~J., 2012, \aap, 548, A86

\bibitem[{{Liseau} {et~al.}(2010){Liseau}, {Eiroa}, {Fedele}, {Augereau},
  {Olofsson}, {Gonz{\'a}lez}, {Maldonado}, {Montesinos}, {Mora}, {Absil},
  {Ardila}, {Barrado}, {Bayo}, {Beichman}, {Bryden}, {Danchi}, {Del Burgo},
  {Ertel}, {Fridlund}, {Heras}, {Krivov}, {Launhardt}, {Lebreton}, {L{\"o}hne},
  {Marshall}, {Meeus}, {M{\"u}ller}, {Pilbratt}, {Roberge}, {Rodmann},
  {Solano}, {Stapelfeldt}, {Th{\'e}bault}, {White}, \& {Wolf}}]{Liseau2010}
{Liseau} R., {Eiroa} C., {Fedele} D., {Augereau} J.-C., {Olofsson} G.,
  {Gonz{\'a}lez} B., {Maldonado} J., {Montesinos} B., {Mora} A., {Absil} O.,
  {Ardila} D., {Barrado} D., {Bayo} A., {Beichman} C.~A., {Bryden} G., {Danchi}
  W.~C., {Del Burgo} C., {Ertel} S., {Fridlund} C.~W.~M., {Heras} A.~M.,
  {Krivov} A.~V., {Launhardt} R., {Lebreton} J., {L{\"o}hne} T., {Marshall}
  J.~P., {Meeus} G., {M{\"u}ller} S., {Pilbratt} G.~L., {Roberge} A., {Rodmann}
  J., {Solano} E., {Stapelfeldt} K.~R., {Th{\'e}bault} P., {White} G.~J.,
  {Wolf} S., 2010, \aap, 518, L132

\bibitem[{{L{\"o}hne} {et~al.}(2012){L{\"o}hne}, {Augereau}, {Ertel},
  {Marshall}, {Eiroa}, {Mora}, {Absil}, {Stapelfeldt}, {Th{\'e}bault}, {Bayo},
  {Del Burgo}, {Danchi}, {Krivov}, {Lebreton}, {Letawe}, {Magain}, {Maldonado},
  {Montesinos}, {Pilbratt}, {White}, \& {Wolf}}]{Lohne2012}
{L{\"o}hne} T., {Augereau} J.-C., {Ertel} S., {Marshall} J.~P., {Eiroa} C.,
  {Mora} A., {Absil} O., {Stapelfeldt} K., {Th{\'e}bault} P., {Bayo} A., {Del
  Burgo} C., {Danchi} W., {Krivov} A.~V., {Lebreton} J., {Letawe} G., {Magain}
  P., {Maldonado} J., {Montesinos} B., {Pilbratt} G.~L., {White} G.~J., {Wolf}
  S., 2012, \aap, 537, A110

\bibitem[{{Marois} {et~al.}(2006){Marois}, {Lafreni{\`e}re}, {Doyon},
  {Macintosh}, \& {Nadeau}}]{Marois2006}
{Marois} C., {Lafreni{\`e}re} D., {Doyon} R., {Macintosh} B., {Nadeau} D.,
  2006, \apj, 641, 556

\bibitem[{{Marois} {et~al.}(2008){Marois}, {Macintosh}, {Barman}, {Zuckerman},
  {Song}, {Patience}, {Lafreni{\`e}re}, \& {Doyon}}]{hr8799detection08}
{Marois} C., {Macintosh} B., {Barman} T., {Zuckerman} B., {Song} I., {Patience}
  J., {Lafreni{\`e}re} D., {Doyon} R., 2008, Science, 322, 1348

\bibitem[{{Marois} {et~al.}(2010){Marois}, {Zuckerman}, {Konopacky},
  {Macintosh}, \& {Barman}}]{4thplanetHR8799}
{Marois} C., {Zuckerman} B., {Konopacky} Q.~M., {Macintosh} B., {Barman} T.,
  2010, \nat, 468, 1080

\bibitem[{{Matthews} {et~al.}(2010){Matthews}, {Sibthorpe}, {Kennedy},
  {Phillips}, {Churcher}, {Duch{\^e}ne}, {Greaves}, {Lestrade}, {Moro-Martin},
  {Wyatt}, {Bastien}, {Biggs}, {Bouvier}, {Butner}, {Dent}, {di Francesco},
  {Eisl{\"o}ffel}, {Graham}, {Harvey}, {Hauschildt}, {Holland}, {Horner},
  {Ibar}, {Ivison}, {Johnstone}, {Kalas}, {Kavelaars}, {Rodriguez}, {Udry},
  {van der Werf}, {Wilner}, \& {Zuckerman}}]{Matthews2010}
{Matthews} B.~C., {Sibthorpe} B., {Kennedy} G., {Phillips} N., {Churcher} L.,
  {Duch{\^e}ne} G., {Greaves} J.~S., {Lestrade} J.-F., {Moro-Martin} A.,
  {Wyatt} M.~C., {Bastien} P., {Biggs} A., {Bouvier} J., {Butner} H.~M., {Dent}
  W.~R.~F., {di Francesco} J., {Eisl{\"o}ffel} J., {Graham} J., {Harvey} P.,
  {Hauschildt} P., {Holland} W.~S., {Horner} J., {Ibar} E., {Ivison} R.~J.,
  {Johnstone} D., {Kalas} P., {Kavelaars} J., {Rodriguez} D., {Udry} S., {van
  der Werf} P., {Wilner} D., {Zuckerman} B., 2010, \aap, 518, L135

\bibitem[{{Mayor} \& {Queloz}(1995)}]{Mayor1995}
{Mayor} M., {Queloz} D., 1995, \nat, 378, 355

\bibitem[{{Mermilliod}(1987)}]{Mermilliod1987}
{Mermilliod} J.-C., 1987, \aaps, 71, 413

\bibitem[{{Moerchen} {et~al.}(2011){Moerchen}, {Churcher}, {Telesco}, {Wyatt},
  {Fisher}, \& {Packham}}]{Moerchen10}
{Moerchen} M.~M., {Churcher} L.~J., {Telesco} C.~M., {Wyatt} M., {Fisher}
  R.~S., {Packham} C., 2011, \aap, 526, A34+

\bibitem[{{Morel} \& {Magnenat}(1978)}]{Morel1978}
{Morel} M., {Magnenat} P., 1978, \aaps, 34, 477

\bibitem[{{Moshir} {et~al.}(1993){Moshir}, {Copan}, {Conrow}, {McCallon},
  {Hacking}, {Gregorich}, {Rohrbach}, {Melnyk}, {Rice}, \&
  {Fullmer}}]{Moshir1993}
{Moshir} M., {Copan} G., {Conrow} T., {McCallon} H., {Hacking} P., {Gregorich}
  D., {Rohrbach} G., {Melnyk} M., {Rice} W., {Fullmer} L., 1993, VizieR Online
  Data Catalog, 2156, 0

\bibitem[{{M{\"u}ller} {et~al.}(2010){M{\"u}ller}, {L{\"o}hne}, \&
  {Krivov}}]{muller}
{M{\"u}ller} S., {L{\"o}hne} T., {Krivov} A.~V., 2010, \apj, 708, 1728

\bibitem[{{Mustill} \& {Wyatt}(2009)}]{alex}
{Mustill} A.~J., {Wyatt} M.~C., 2009, \mnras, 399, 1403

\bibitem[{{Ott}(2010)}]{Ott2010}
{Ott} S., 2010, 434, 139

\bibitem[{{Perryman} \& {ESA}(1997)}]{Perryman1997}
{Perryman} M.~A.~C., {ESA}, 1997, 1200

\bibitem[{{Poglitsch} {et~al.}(2010){Poglitsch}, {Waelkens}, {Geis},
  {Feuchtgruber}, {Vandenbussche}, {Rodriguez}, {Krause}, {Renotte}, {van
  Hoof}, {Saraceno}, {Cepa}, {Kerschbaum}, {Agn{\`e}se}, {Ali}, {Altieri},
  {Andreani}, {Augueres}, {Balog}, {Barl}, {Bauer}, {Belbachir}, {Benedettini},
  {Billot}, {Boulade}, {Bischof}, {Blommaert}, {Callut}, {Cara}, {Cerulli},
  {Cesarsky}, {Contursi}, {Creten}, {De Meester}, {Doublier}, {Doumayrou},
  {Duband}, {Exter}, {Genzel}, {Gillis}, {Gr{\"o}zinger}, {Henning},
  {Herreros}, {Huygen}, {Inguscio}, {Jakob}, {Jamar}, {Jean}, {de Jong},
  {Katterloher}, {Kiss}, {Klaas}, {Lemke}, {Lutz}, {Madden}, {Marquet},
  {Martignac}, {Mazy}, {Merken}, {Montfort}, {Morbidelli}, {M{\"u}ller},
  {Nielbock}, {Okumura}, {Orfei}, {Ottensamer}, {Pezzuto}, {Popesso},
  {Putzeys}, {Regibo}, {Reveret}, {Royer}, {Sauvage}, {Schreiber}, {Stegmaier},
  {Schmitt}, {Schubert}, {Sturm}, {Thiel}, {Tofani}, {Vavrek}, {Wetzstein},
  {Wieprecht}, \& {Wiezorrek}}]{Poglitsch2010}
{Poglitsch} A., {Waelkens} C., {Geis} N., {Feuchtgruber} H., {Vandenbussche}
  B., {Rodriguez} L., {Krause} O., {Renotte} E., {van Hoof} C., {Saraceno} P.,
  {Cepa} J., {Kerschbaum} F., {Agn{\`e}se} P., {Ali} B., {Altieri} B.,
  {Andreani} P., {Augueres} J.-L., {Balog} Z., {Barl} L., {Bauer} O.~H.,
  {Belbachir} N., {Benedettini} M., {Billot} N., {Boulade} O., {Bischof} H.,
  {Blommaert} J., {Callut} E., {Cara} C., {Cerulli} R., {Cesarsky} D.,
  {Contursi} A., {Creten} Y., {De Meester} W., {Doublier} V., {Doumayrou} E.,
  {Duband} L., {Exter} K., {Genzel} R., {Gillis} J.-M., {Gr{\"o}zinger} U.,
  {Henning} T., {Herreros} J., {Huygen} R., {Inguscio} M., {Jakob} G., {Jamar}
  C., {Jean} C., {de Jong} J., {Katterloher} R., {Kiss} C., {Klaas} U., {Lemke}
  D., {Lutz} D., {Madden} S., {Marquet} B., {Martignac} J., {Mazy} A., {Merken}
  P., {Montfort} F., {Morbidelli} L., {M{\"u}ller} T., {Nielbock} M., {Okumura}
  K., {Orfei} R., {Ottensamer} R., {Pezzuto} S., {Popesso} P., {Putzeys} J.,
  {Regibo} S., {Reveret} V., {Royer} P., {Sauvage} M., {Schreiber} J.,
  {Stegmaier} J., {Schmitt} D., {Schubert} J., {Sturm} E., {Thiel} M., {Tofani}
  G., {Vavrek} R., {Wetzstein} M., {Wieprecht} E., {Wiezorrek} E., 2010, \aap,
  518, L2

\bibitem[{{Sato} {et~al.}(2010){Sato}, {Omiya}, {Liu}, {Harakawa}, {Izumiura},
  {Kambe}, {Toyota}, {Murata}, {Lee}, {Masuda}, {Takeda}, {Yoshida}, {Itoh},
  {Ando}, {Kokubo}, {Ida}, {Zhao}, \& {Han}}]{Sato2010}
{Sato} B., {Omiya} M., {Liu} Y., {Harakawa} H., {Izumiura} H., {Kambe} E.,
  {Toyota} E., {Murata} D., {Lee} B.-C., {Masuda} S., {Takeda} Y., {Yoshida}
  M., {Itoh} Y., {Ando} H., {Kokubo} E., {Ida} S., {Zhao} G., {Han} I., 2010,
  \pasj, 62, 1063

\bibitem[{{Su} {et~al.}(2005){Su}, {Rieke}, {Misselt}, {Stansberry},
  {Moro-Martin}, {Stapelfeldt}, {Werner}, {Trilling}, {Bendo}, {Gordon},
  {Hines}, {Wyatt}, {Holland}, {Marengo}, {Megeath}, \& {Fazio}}]{vegasu05}
{Su} K.~Y.~L., {Rieke} G.~H., {Misselt} K.~A., {Stansberry} J.~A.,
  {Moro-Martin} A., {Stapelfeldt} K.~R., {Werner} M.~W., {Trilling} D.~E.,
  {Bendo} G.~J., {Gordon} K.~D., {Hines} D.~C., {Wyatt} M.~C., {Holland} W.~S.,
  {Marengo} M., {Megeath} S.~T., {Fazio} G.~G., 2005, \apj, 628, 487

\bibitem[{{Su} {et~al.}(2009){Su}, {Rieke}, {Stapelfeldt}, {Malhotra},
  {Bryden}, {Smith}, {Misselt}, {Moro-Martin}, \& {Williams}}]{Hr8799su}
{Su} K.~Y.~L., {Rieke} G.~H., {Stapelfeldt} K.~R., {Malhotra} R., {Bryden} G.,
  {Smith} P.~S., {Misselt} K.~A., {Moro-Martin} A., {Williams} J.~P., 2009,
  \apj, 705, 314

\bibitem[{{Wisdom}(1980)}]{Wisdom1980}
{Wisdom} J., 1980, \aj, 85, 1122

\bibitem[{{Wizinowich} {et~al.}(2000){Wizinowich}, {Acton}, {Shelton},
  {Stomski}, {Gathright}, {Ho}, {Lupton}, {Tsubota}, {Lai}, {Max}, {Brase},
  {An}, {Avicola}, {Olivier}, {Gavel}, {Macintosh}, {Ghez}, \&
  {Larkin}}]{Wizinowich2000}
{Wizinowich} P., {Acton} D.~S., {Shelton} C., {Stomski} P., {Gathright} J.,
  {Ho} K., {Lupton} W., {Tsubota} K., {Lai} O., {Max} C., {Brase} J., {An} J.,
  {Avicola} K., {Olivier} S., {Gavel} D., {Macintosh} B., {Ghez} A., {Larkin}
  J., 2000, \pasp, 112, 315

\bibitem[{{Wolszczan} \& {Frail}(1992)}]{pulsarplanets}
{Wolszczan} A., {Frail} D.~A., 1992, \nat, 355, 145

\bibitem[{{Wyatt}(2008)}]{wyattreview}
{Wyatt} M.~C., 2008, \araa, 46, 339

\bibitem[{{Wyatt} {et~al.}(1999){Wyatt}, {Dermott}, {Telesco}, {Fisher},
  {Grogan}, {Holmes}, \& {Pi{\~n}a}}]{Wyatt99}
{Wyatt} M.~C., {Dermott} S.~F., {Telesco} C.~M., {Fisher} R.~S., {Grogan} K.,
  {Holmes} E.~K., {Pi{\~n}a} R.~K., 1999, \apj, 527, 918

\bibitem[{{Wyatt} {et~al.}(2012){Wyatt}, {Kennedy}, {Sibthorpe},
  {Moro-Mart{\'{\i}}n}, {Lestrade}, {Ivison}, {Matthews}, {Udry}, {Greaves},
  {Kalas}, {Lawler}, {Su}, {Rieke}, {Booth}, {Bryden}, {Horner}, {Kavelaars},
  \& {Wilner}}]{Wyatt61Vir}
{Wyatt} M.~C., {Kennedy} G., {Sibthorpe} B., {Moro-Mart{\'{\i}}n} A.,
  {Lestrade} J.-F., {Ivison} R.~J., {Matthews} B., {Udry} S., {Greaves} J.~S.,
  {Kalas} P., {Lawler} S., {Su} K.~Y.~L., {Rieke} G.~H., {Booth} M., {Bryden}
  G., {Horner} J., {Kavelaars} J.~J., {Wilner} D., 2012, \mnras, 424, 1206

\bibitem[{{Wyatt} {et~al.}(2007{\natexlab{a}}){Wyatt}, {Smith}, {Greaves},
  {Beichman}, {Bryden}, \& {Lisse}}]{Wyatt07hot}
{Wyatt} M.~C., {Smith} R., {Greaves} J.~S., {Beichman} C.~A., {Bryden} G.,
  {Lisse} C.~M., 2007{\natexlab{a}}, \apj, 658, 569

\bibitem[{{Wyatt} {et~al.}(2007{\natexlab{b}}){Wyatt}, {Smith}, {Su}, {Rieke},
  {Greaves}, {Beichman}, \& {Bryden}}]{wyatt07}
{Wyatt} M.~C., {Smith} R., {Su} K.~Y.~L., {Rieke} G.~H., {Greaves} J.~S.,
  {Beichman} C.~A., {Bryden} G., 2007{\natexlab{b}}, \apj, 663, 365

\bibitem[{{Zuckerman} {et~al.}(1995){Zuckerman}, {Kim}, \&
  {Liu}}]{zuckermankim95}
{Zuckerman} B., {Kim} S.~S., {Liu} T., 1995, \apjl, 446, L79+

\end{thebibliography}

\end{document}